\documentclass[12pt,latexsym,amssymb]{article}

\usepackage{epsfig,multicol,multirow,cite}
\usepackage{amsmath,subfigure,latexsym,amssymb}

\def\lsi{\raise0.3ex\hbox{$<$\kern-0.75em\raise-1.1ex\hbox{$\sim$}}}
\def\gsi{\raise0.3ex\hbox{$>$\kern-0.75em\raise-1.1ex\hbox{$\sim$}}}

\def\backder{\raise1.4ex\hbox{$\leftarrow$\kern-0.75em\raise-1.4ex\hbox{$\partial$}}}

\newcommand{\ZZ}{\mathbb{Z}}
\newcommand{\RR}{\mathbb{R}}

\newcommand{\NN}{\mathbb{N}}
\newcommand{\R}{{\kern+.25em\sf{R}\kern-.78em\sf{I} \kern+.78em\kern-.25em}}

\newcommand{\la}{\langle}
\newcommand{\ra}{\rangle}

\newcommand{\ri}{{\rm i}}

\newcommand{\nn}{\nonumber}

\newcommand{\be}{\begin{equation}}
\newcommand{\ee}{\end{equation}}
\newcommand{\bea}{\begin{eqnarray}}
\newcommand{\eea}{\end{eqnarray}}

\newcommand{\vp}{\varphi}

\newcommand{\gtapprox}{\raisebox{-0.5ex}{$\,\stackrel{>}{\scriptstyle\sim}\,$}}
\newcommand{\ltapprox}{\raisebox{-0.5ex}{$\,\stackrel{<}{\scriptstyle\sim}\,$}}

\makeatletter
\@addtoreset{equation}{section}
\makeatother

\begin{document}

\begin{center}

{\large\bf{Berezinski\u{\i}-Kosterlitz-Thouless Transition and the Haldane 
\vspace*{3mm} \\
Conjecture: Highlights of the Physics Nobel Prize 2016}} \vspace*{8mm} \\

Wolfgang Bietenholz$^{\rm \, a,b}$ and Urs Gerber$^{\rm \, a,c}$\\
\ \\
$^{\rm a}$ Instituto de Ciencias Nucleares \\
Universidad Nacional Aut\'{o}noma de M\'{e}xico\\
A.P.\ 70-543, C.P.\ 04510 Ciudad de M\'{e}xico, Mexico\\
\ \vspace*{-2mm} \\
$^{\rm b}$ Albert Einstein Center for Fundamental Physics \\
Institute for Theoretical Physics \\ 
University of Bern, Sidlerstrasse 5, CH-3012 Bern, Switzerland \\
\ \vspace*{-2mm} \\
$^{\rm c}$ Instituto de F\'{\i}sica y Matem\'{a}ticas \\
Universidad Michoacana de San Nicol\'{a}s de Hidalgo, Edificio C-3\\
Apdo.\ Postal 2-82, C.P.\ 58040, Morelia, Michoac\'{a}n, Mexico 

\end{center}

\vspace*{5mm}

\noindent
The 2016 Physics Nobel Prize honors a variety of discoveries 
related to topological phases and phase transitions. Here
we sketch two exciting facets: the groundbreaking works by
John Kosterlitz and David Thouless on phase transitions of infinite 
order, and by Duncan Haldane on the energy gaps in quantum spin 
chains. These insights came as surprises in the 1970s and 1980s, 
respectively, and they have both initiated new fields of 
research in theoretical and experimental physics.\\

\noindent
{\footnotesize PACS: General theory of phase transitions, 64.60.Bd;
Statistical mechanics of model systems, 64.60.De;
General theory of critical region behavior, 64.60.fd;
Equilibrium properties near critical points, critical exponents, 64.60.F-}

\section{Classical Spin Models}

When we hear the word ``spin'' we usually think of Quantum Mechanics,
where particles are endowed with an internal degree of freedom, which 
manifests itself like an angular momentum.
So what does a ``classical spin'' mean?

It is much simpler: it is just a vector (or multi-scalar)
$\vec e$, say with $N$ components; here we assume them to be real,
\be
\vec e = \left( \begin{array}{c} e^{(1)} \\ \cdot \\ \cdot \\
e^{(N)} \end{array} \right) \in \RR^{N} \ .
\ee
Models which deal with such {\em classical spin fields} are often
formulated on a lattice (or grid), such that a spin $\vec e_{x}$ is 
attached to each lattice site $x$. 
In solid state physics, $\vec e_{x}$ might represent a collective 
spin of some crystal cell. If it is composed of many 
quantum spins, it appears classical \cite{Ma}.

If the spin direction is fixed at each site $x$, we obtain
a {\em configuration,} which we denote as $[ \vec e \, ]$.

In a number of very popular models, the length of each spin variable 
is normalized to $|\vec e_{x}| =  1, \ \forall x$. Then the
spin field maps the sites onto a unit sphere in the $N$-dimensional
spin space, $x \to S^{N-1}$. 
We are going to refer to this setting, and (for simplicity) to a 
lattice of unit spacing,  with sites $x \in \ZZ^{d}$ in $d$ dimensions.

To define a model, we still need to specify a {\em Hamilton function}
${\cal H} [\vec e \, ]$ (no operator), which fixes the energy of any
possible spin configuration. Its standard form reads
\be \label{Hami}
{\cal H}[\vec e \, ] = J \sum_{ \la xy \ra} (1 - \vec e_{x} 
\cdot \vec e_{y}) - \vec H \cdot \sum_{x} \vec e_{x} \ ,
\ee
where the symbol $\la xy \ra$ denotes nearest neighbor sites.
$J$ is a coupling constant, and we see that $J>0$ describes
a {\em ferromagnetic} behavior: (approximately) parallel spins
are favored, since they minimize the energy.\footnote{Vice versa, 
$J<0$ describes anti-ferromagnets, which also occur in models 
(cf.\ Section 3) and in Nature, {\it e.g.}\ Cr, Mn, 
Fe$_{2}$O$_{3}$ and NiS$_{2}$.\label{affoot}} $\vec H$ is an
 external ``magnetic field'' (an ``ordering field'', in a generalized
sense), which may or may not be included;\footnote{In field theory
one usually deals with ``source fields'', which correspond to a space
dependent external field of this kind, {\it i.e.}\ to a term
$ \sum_{x}  \vec H_{x} \cdot \vec e_{x}\,$.}
its presence favors spin orientations in the direction of $\vec H$.

Thus we arrive at a set of highly prominent models in statistical 
mechanics, depending on the spin dimension $N $: \\

\begin{tabular}{c c c}
$N=1$ & $e_{x} \in \{ -1, +1 \}$ & Ising model \\
$N=2$ & $ \vec e_{x}^{\rm ~T} = (\cos \vp_{x}, \sin \vp_{x})$
& XY model \\
$N=3$ &
$\vec e_{x}^{\rm ~T} = (\sin \theta_{x} \cos \vp_{x}, \sin \theta_{x} \sin \vp_{x},
\cos \theta_{x})$
& Heisenberg model
\end{tabular} \\

\noindent where $\vp_{x}, \, \theta_{x} \in \RR$. 
These models are discussed in numerous text books,
such as Refs.\ \cite{Ma,critphen,ZJ}.

Although it might seem ridiculously simple, the {\em Ising model}
is incredibly successful in describing a whole host of physical 
phenomena. The {\em XY model} will be addressed in Section 2; its
best application is to model superfluid helium. The {\em Heisenberg 
model} captures actual ferromagnets, like iron, cobalt and nickel.
Section 3.1 refers to its 2d version,  which is also
a toy model for Quantum Chromodynamics (QCD), since
it shares fundamental properties like asymptotic freedom,
topological sectors, and a dynamically generated mass gap.
The {\em large $N$ limit} also attracts attention, since it leads
to simplifications, which enable analytical calculations, 
see {\it e.g.}\ Ref.\ \cite{ZJ}.

They are also called non-linear $\sigma$-models, or
O($N$) models, since --- with the Hamilton function 
(\ref{Hami}) at $\vec H = \vec 0$ --- they have
a global O($N$) symmetry (or $Z(2)$ symmetry in case of the Ising 
model): the energy remains invariant if we perform
the same rotation on all spins, $\vec e_{x} \to \Omega \, \vec e_{x},
\ \Omega \in {\rm O}(N)$.\footnote{The O(4) model is of 
interest as well, in particular due to the local isomorphy
${\rm O}(4) \sim {\rm SU}(2) \otimes {\rm SU}(2)$.
The latter is the flavor chiral symmetry of QCD
with two massless flavors. Here the magnetic field corresponds to
a small mass of the quark flavors $u$ and $d$, which breaks 
the symmetry down to ${\rm O}(3) \sim {\rm SU}(2)$.}

If the system has temperature $T$, the probability for a configuration 
$[\vec e \, ]$ is given by\footnote{We express the temperature in 
units of the Boltzmann constant $k_{\rm B}$, which amounts to 
setting $k_{\rm B}=1$ throughout this article.}
\be  \label{prob}
p[\vec e \, ] = \frac{1}{Z} \ e^{- {\cal H}[\vec e \, ] / T } \ , \quad
{\rm with} \quad Z = \sum_{[\vec e \, ]} 
e^{- {\cal H}[\vec e \, ] / T } = e^{-F/T} \ .
\ee
The {\em partition function} $Z$ is obtained by summing
(or integrating) over {\em all} possible configurations,\footnote{In
$N\geq 2$ the number of configurations is infinite. For the Ising
model in a lattice volume $V$, {\it i.e.}\ with $V$ lattice sites,
their number is $2^{V}$. Even for a modest volume, say a
$32 \times 32$ lattice, this is a huge number of $O(10^{308})$, so 
straight summation is not feasible, not even with supercomputers.
Hence to compute expectation values (see below) one resorts to
{\em importance sampling} by means of Monte Carlo simulations. For 
a text book and a recent introductory review, see Refs.\ 
\cite{hadphys}.\label{lattice}}
and $F = - T \ln Z$ is the {\em free energy}.

\begin{figure}[h!]
\begin{center}
\includegraphics[width=0.3\textwidth,angle=0]{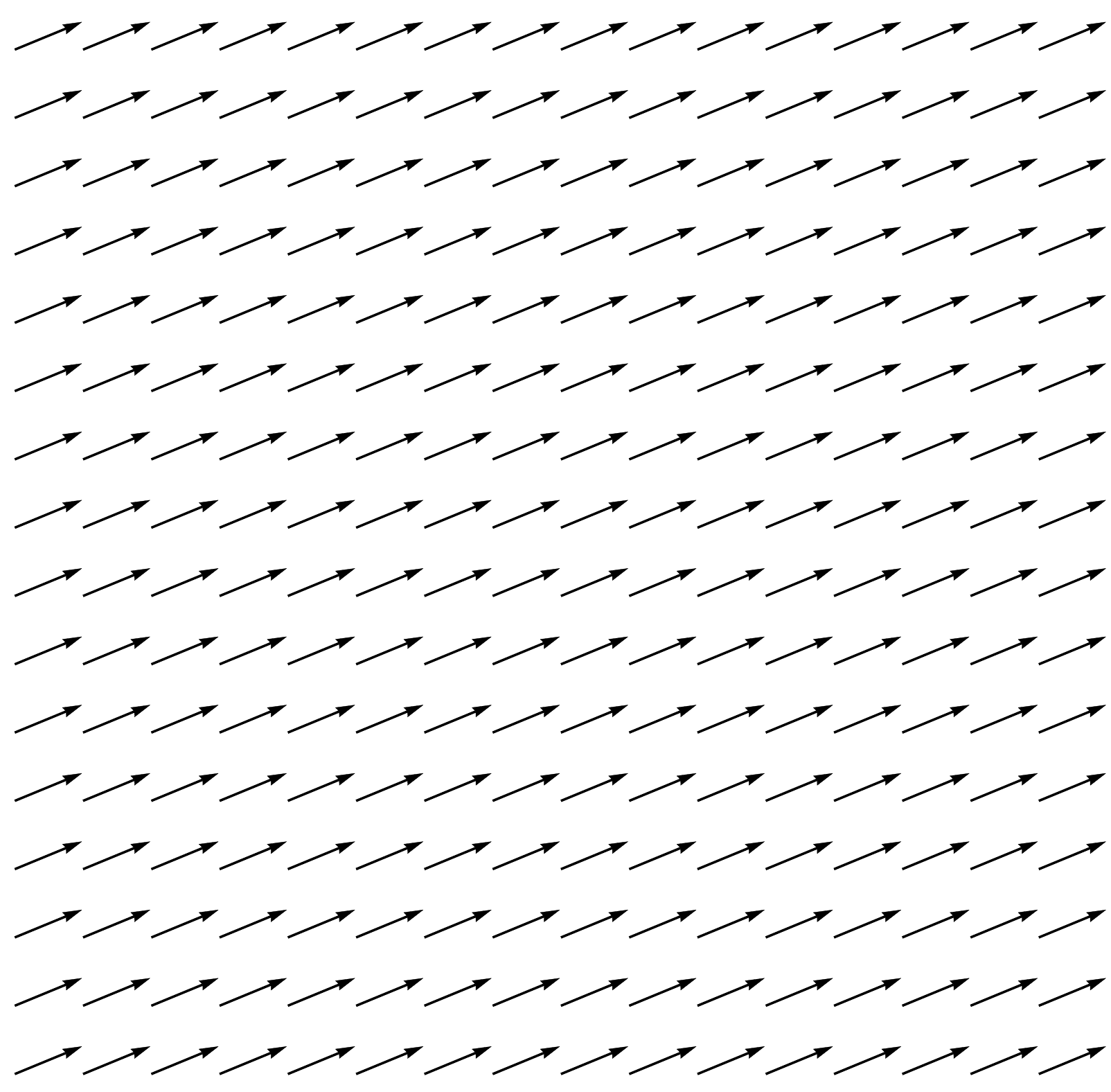}
\hspace{1cm}
\includegraphics[width=0.3\textwidth,angle=0]{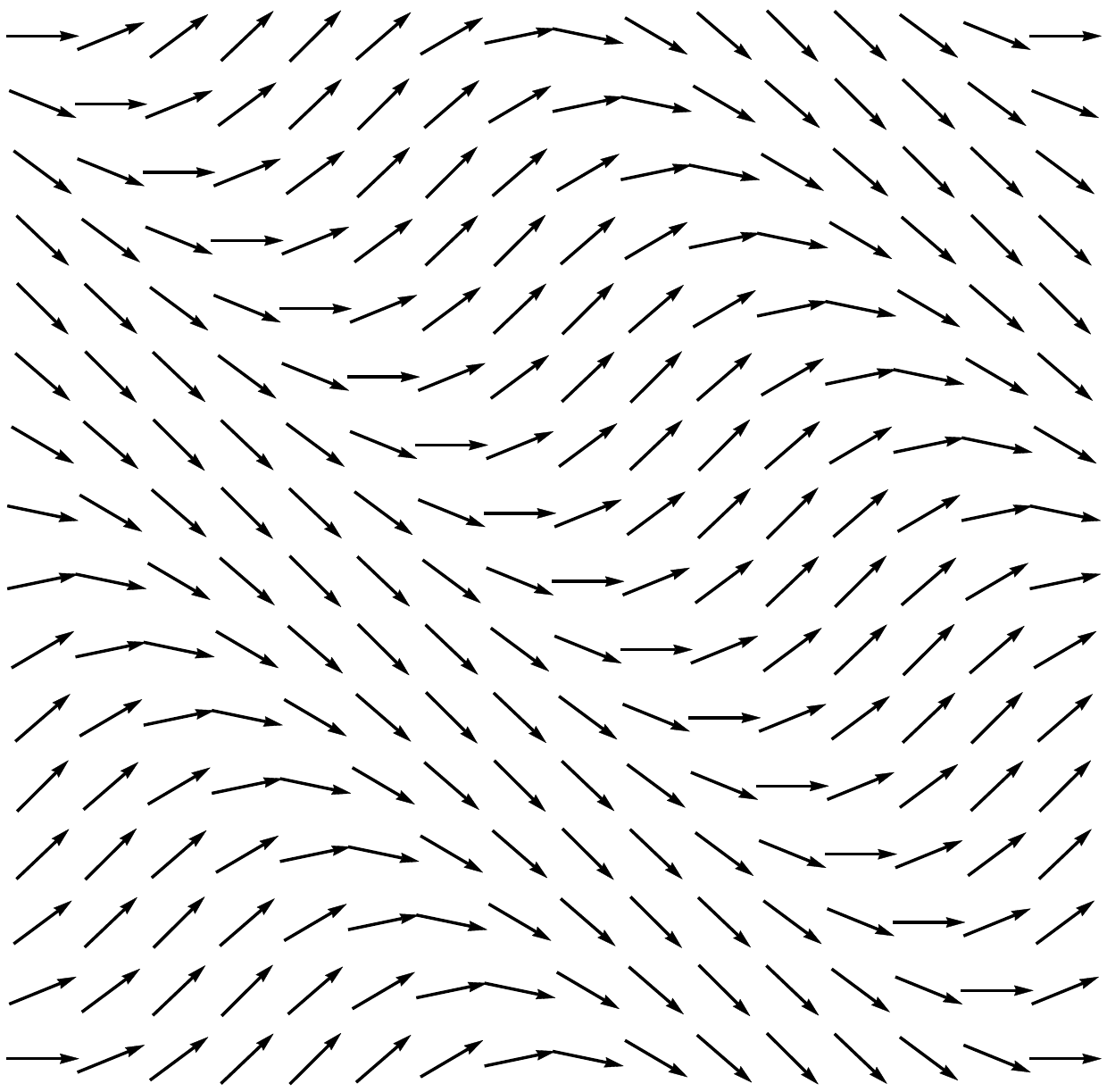}
\caption{Examples for a uniform configuration of minimal energy (left) and 
for a non-uniform configuration of higher energy (right), in the 2d XY model.}
\label{novort}
\end{center}
\vspace*{-5mm}
\end{figure}

For $J> 0$ the uniform configurations are most probable, since 
they have the minimal energy $-V H$ (where $H = |\vec H|$). 
An example is shown in Fig.\ \ref{novort} (left), and in the limit 
$T \to 0$ the system will take such a uniform configuration.
For increasing $T$, fluctuating configurations --- like the one
in Fig.\  \ref{novort} (right) --- gain more importance. 
They carry higher energy, so the exponential $\exp (- {\cal H}[\vec e \, ]/T)$ 
suppresses them (they are less suppressed for increasing $T$). On the other 
hand, there are {\em many} of them, and the combinatorial factor is relevant 
too. This is the {\em entropy effect,} which also matters for their 
impact, and which plays a key r\^{o}le in Section 2.

\subsection{$n$-point functions and phase transitions}

What does it mean to have an ``impact''? What physical
quantities are affected? In exact analogy to field theory,
the physical terms are expectation values of some 
products of spins; if they involve $n$ factors, they are
called {\em $n$-point functions.} \\

The most important observable is the 2-point function,
or {\em correlation function,}
\be  \label{corre}
\la \vec e_{x} \cdot \vec e_{y} \ra = \frac{1}{Z} \ \sum_{[\vec e \, ]} 
\vec e_{x} \cdot \vec e_{y} \ e^{- {\cal H}[\vec e \, ] / T } \ .
\ee
One often focuses on its ``connected part'', which --- in most 
cases --- decays exponentially in the distance $|x-y|$,
\be  \label{expdecay}
\la \vec e_{x} \cdot \vec e_{y} \ra_{\rm con} 
= \la \vec e_{x} \cdot \vec e_{y} \ra
-  \la \vec e_{x} \ra \cdot \la \vec e_{y} \ra 
= \la \vec e_{x} \cdot \vec e_{y} \ra
- \la \vec e \, \ra^{2}
\propto e^{-| x - y| / \xi} \ .
\ee
With the Hamilton function (\ref{Hami}) the system is lattice
translation invariant, so the 1-point function $\la \vec e_{x} \ra$ 
does not depend on the site $x$, and we can just write 
$\la \vec e \, \ra$,
\be  \label{1point}
\la \vec e_{x} \ra = \frac{1}{Z} \ \sum_{[\vec e \, ]} 
\vec e_{x} \ e^{- {\cal H}[\vec e \, ] / T } \  = \la \vec e \, \ra \ .
\ee

The decay rate of $\la \vec e_{x} \cdot \vec e_{y} \ra_{\rm con}$
is given by the {\em correlation length} $\xi$,
which serves as {\em the} scale of the system: any dimensional quantity
is considered ``large'' or ``small'' based on its comparison with
(the suitable power of) $\xi$. Regarding the energy spectrum,
$\xi$ represents the inverse
{\em energy gap,} $1/\xi = E_{1} - E_{0}$.
In quantum field theory, this is just the {\em mass} of the
particle, which emerges by the minimal (quantized) excitations of 
the field under consideration.

The phase transitions that we are interested in are of order 2 or 
higher, and they are characterized by the property that $\xi$ 
{\em diverges.} In a phase diagram, with axes like $T$ and $H$,
this happens in a {\em critical point,}\footnote{Phase transitions 
of first order are more frequent, and they do not correspond to 
a critical point, but we won't discuss them.}
in particular at a {\em critical temperature} $T_{\rm c}$.
The way how $\xi$ diverges in the vicinity of a critical point
defines the {\em critical exponent $\nu$,}
\be  \label{criteta}
\lim_{T \to T_{\rm c}} \ \xi \propto (T - T_{\rm c})^{-\nu} \ ,
\ee
where we assume the same power regardless whether $T_{\rm c}$ 
is approached from above or from below (which usually holds).
There are a number of critical exponents, which characterize
the system close to a critical point; we will
see further examples below.

In the limit $\xi \to \infty$, the spacing between the lattice 
points becomes insignificant (it is negligible compared to $\xi$), 
so this is the {\em continuum limit}.
This is why the vicinity of a critical point is so much of interest.

Intuitively it is clear that high temperature gives importance
to ``wild fluctuations'', which hamper long-distance correlations,
inducing a short $\xi$. 
So does $\xi$ diverge only in the limit $T\to 0 \,$?
This is indeed the case for the 1d Ising model \cite{Ising}.
It does not have an actual transition (with phases on both sides),
and the model is considered uninteresting. However, the Ising 
model does have a finite critical temperature in dimension $d=2$
\cite{2dIsing} or higher, and the same applies to $N>1$. \\

The simplest observable is the 1-point function, or {\em condensate},
given in eq.\ (\ref{1point}), which also defines the {\em magnetization} 
$M$ (in some lattice volume $V$),
\be
\vec m [\vec e \, ] = \sum_{x} \vec e_{x} \ , \quad
M = | \la \vec m \ra | = V | \la \vec e \, \ra | \ .
\ee
$M>0$ indicates that the O($N$) symmetry is broken.
An external field $H >0$ causes its explicit breaking. 
If we start with an external field 
(in {\em infinite} volume, $V \to \infty$)
and gradually turn it off, the destiny of the system 
depends on the temperature:
\begin{itemize}

\item At low $T$, the system keeps a dominant orientation
in the direction of $\vec H$, at $T \to 0$ it will pick the 
corresponding uniform configuration. This is known as 
``spontaneous symmetry breaking'', it reduces the symmetry
group to O($N-1$).

\item At high $T$, the system allows for wild fluctuations, and
after turning off $\vec H$ it hardly ``remembers'' its
direction. In this case, the O($N$) symmetry is restored,
since the dominant contributions to an expectation value are
due to configurations without such a preferred orientation.

\end{itemize}

Thus the magnetization $M$ discriminates the scenarios where the 
O($N$) symmetry is broken ($M>0$, {\em order}) or intact ($M \simeq 0$, 
{\em disorder}). Therefore it is an {\em order parameter:} it is finite 
(it vanishes) below (above) the critical temperature $T_{\rm c}$, which 
is also called Curie temperature.
The way how it converges to $0$, as $T$ approaches $T_{\rm c}$ 
from below, defines another critical exponent $\beta$,
\be
\lim_{T \nearrow T_{\rm c}} \, M \propto (T_{\rm c} -T)^{\beta} \ .
\ee
The follow-up example is the critical exponent $\gamma$, which
characterizes the divergence of the magnetic susceptibility 
$\chi_{\rm m}$, at a temperature $T$ close to $T_{\rm c}$,
\be  \label{gamma}
\chi_{\rm m} 
= \frac{1}{V} \Big( \la \vec m^{\, 2} \ra -  \la \vec m \ra^{2} \Big)
\propto |T - T_{\rm c}|^{-\gamma} \ .
\ee
As in the case of $\nu$, also the exponent $\gamma$ is usually
the same for $T \gtapprox T_{\rm c}$ and for $T \ltapprox T_{\rm c}$.

There are classes of systems, which may look quite different, 
but which share the same critical behavior; we say 
that they belong to the same {\em universality class.}
In particular the (dimensionless) critical exponents coincide within
a universality class. The enormous success of the Ising model is
due to the fact that there are many models --- and real systems ---
in the same universality class, so the Ising model captures their 
behavior next to a continuum limit.

\section{Berezinski\u{\i}-Kosterlitz-Thouless Transition \\
in the 2d XY Model}

This section deals with the 2d XY model, which is among the classical
spin models introduced in Section 1. We can imagine a 2d square 
lattice, where each site $x = (x_{1},x_{2}), \ x_{\mu} \in \ZZ$, carries
a ``watch hand'' $\vec e_{x}$, like an arrow from the origin to 
some point on a unit circle. These arrows are parameterizable by an 
angle $\vp_{x}$, $\vec e_{x} = (\cos \vp_{x}, \sin \vp_{x})$, as we 
mentioned before.

We formulate the angular difference between two spins as
\be
\Delta \vp_{x,y} = (\vp_{y} - \vp_{x}) ~{\rm mod} ~ 2 \pi \in
(-\pi , \pi] \ ,
\ee
{\it i.e.}\ the modulo operation acts such that it picks
the minimal absolute value.

Now let us consider one {\em plaquette,} {\it i.e.}\ one elementary 
square of the lattice with corners $x$, $x +\hat 1$, $x +\hat 2$, 
$x +\hat 1 +\hat 2$, where $\hat \mu$ is a unit vector in 
$\mu$-direction.
For a given configuration, each plaquette has a {\em vortex number} 
$v_{x}$,
\be
v_{x} = \frac{1}{2\pi} \Big( \Delta \vp_{x, x +\hat 1} +
\Delta \vp_{x + \hat 1 , x +\hat 1 + \hat 2} 
+ \Delta \vp_{ x +\hat 1 + \hat 2, x + \hat 2} 
+ \Delta \vp_{ x +\hat 2, x} \Big) \in \{ -1, 0, +1 \} \ .
\ee
If the configuration is smooth (close to uniform) in the range
of this plaquette, we expect $v_{x}=0$.
In case of sizable angular differences 
$|\Delta \vp_{x, x \pm \hat \mu}|$, however, we might encounter a 
{\em topological defect:} this could be a {\em vortex}, which 
we denote as V, or an {\em anti-vortex}, AV. They correspond to
vortex number $+1$ and $-1$, respectively,\\

\begin{center}
\begin{tabular}{ccc}
vortex & V & $v_{x} = +1$ \ \\
anti-vortex & AV & $v_{x} = -1 .$
\end{tabular} \\
\end{center}

\begin{figure}[h!]
\begin{center}
\includegraphics[width=0.3\textwidth,angle=0]{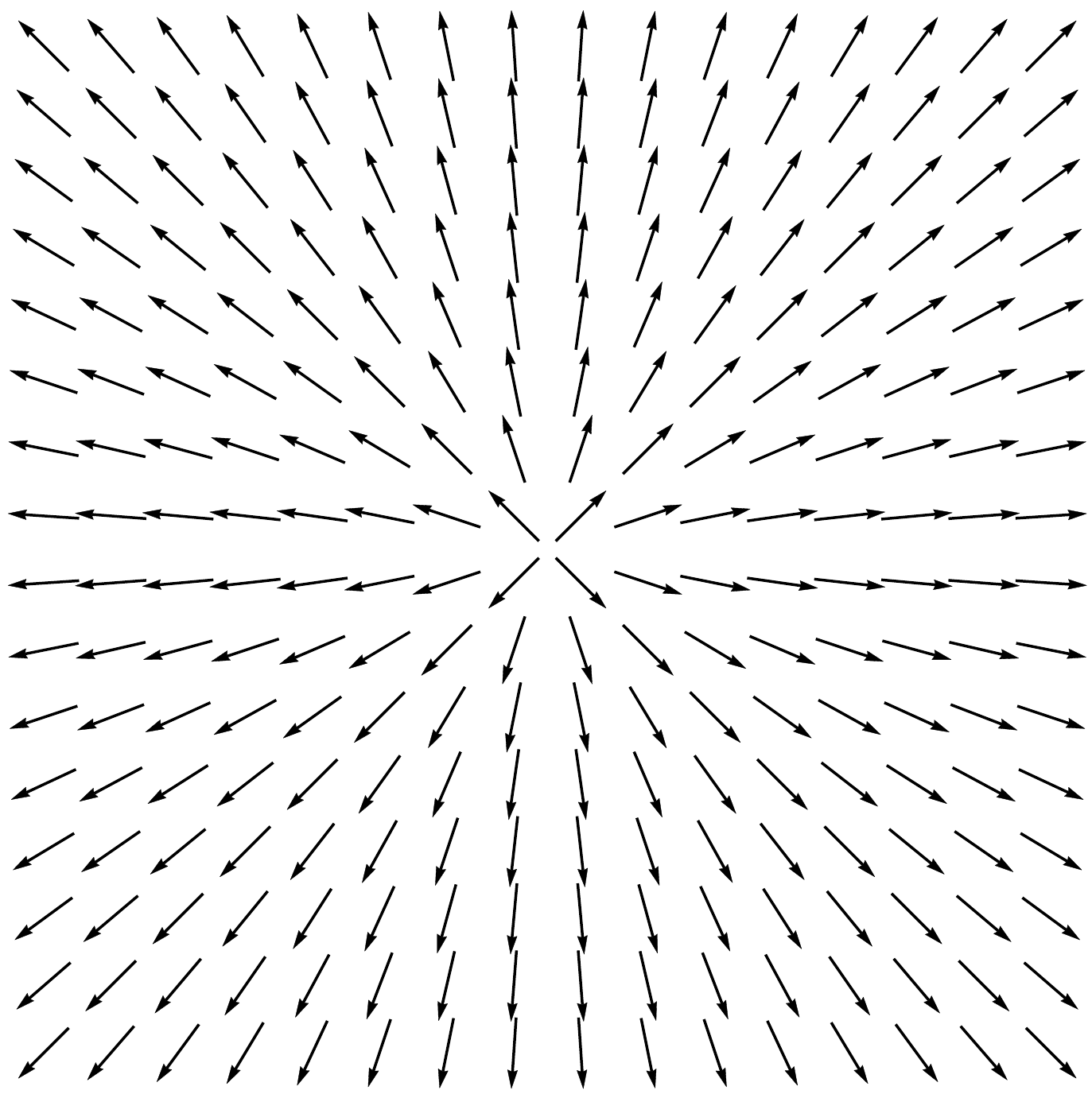}
\hspace{1cm}
\includegraphics[width=0.3\textwidth,angle=0]{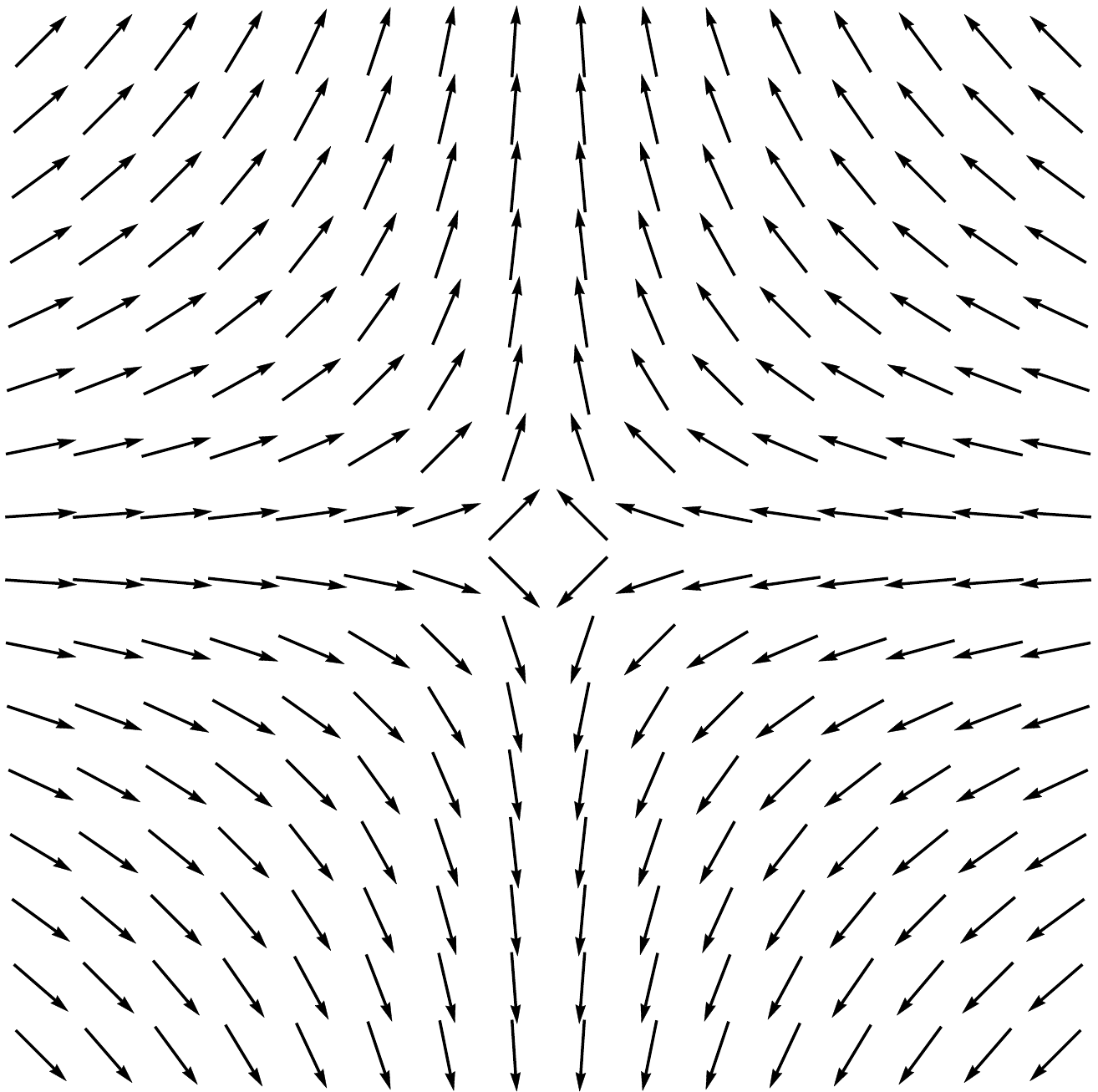}
\caption{Examples for configurations with one vortex V (left), and 
with one anti-vortex AV (right), in the 2d XY model.}
\label{vort}
\end{center}
\vspace*{-5mm}
\end{figure}
\noindent
Examples for a configuration with one V or one AV are shown in 
Fig.\ \ref{vort}. On the other hand, the configurations 
in Fig.\ \ref{novort} do not contain any topological defects.

In numerical studies, we have to deal with a finite lattice volume $V$,
and we usually implement periodic boundary conditions in both directions;
this provides lattice translation invariance. Then the volume
represents a torus, and the total vorticity vanishes, $\sum_{x} v_{x} = 0$, 
due to Stokes' Theorem. So the number of vortices must be equal to the 
number of anti-vortices, $n_{\rm V} = n_{\rm AV}$, and the configurations
of Fig.\ \ref{vort} are actually incompatible with periodic boundaries.

In fact, the global system does not have topological sectors, since its 
homotopy group is trivial, $\Pi_{2}(S^{1}) = \{ 0\}$. Nevertheless, 
the {\em local} topological defects V and AV 
are the crucial degrees of freedom for its phase transition.

\subsection{First look} 

A first look suggests the following picture:

\begin{itemize}

\item The presence of many V and AV, {\it i.e.}\ a {\em high 
vorticity density} 
$$
\rho = \la n_{\rm V} + n_{\rm AV} \ra/V = 2 \la n_{\rm V} \ra/V \ ,
$$
means that strong 
fluctuations are powerful, and they destroy the long-range correlations. 
Hence the corresponding smooth configurations are suppressed,
the correlation function $\la \vec e_{x} \cdot \vec e_{y}\ra$ decays 
rapidly, as in relation (\ref{expdecay}),
and we obtain a correlation length $\xi$ of a few lattice spacings. 
Due to the interpretation of $1/\xi$ as a mass, this is called the 
{\em massive phase.}

\item On the other hand, for a {\em low vorticity density}, $\rho \ll 1$,
long-range correlation dominates. It is not disturbed significantly by the
few V and AV that are floating around, and we are in the {\em massless 
phase}, where $\xi = \infty$. Here the correlation function 
$\la \vec e_{x} \cdot \vec e_{y}\ra$ does not decay exponentially, 
but only with some negative power of $|x-y|$.
The system is {\em conformal}, {\it i.e.}\ scale-invariant.

\end{itemize}

If we start from low temperature and increase $T$ gradually, this
gives more importance to ``rough'' rather than smooth configurations 
--- they are far from uniform, with strong fluctuations. This increases 
the vorticity density $\rho$, and at the critical temperature 
$\rho$ is large enough to mess up the long-range correlations,
so the system enters its massive phase.

To make this point more explicit, we estimate the energy that it 
takes to implement one V or one AV in an otherwise smooth configuration.
We do so in a simplified scheme of a quasi-continuous plane: 
close to the transition this can be justified, since
$\xi$ (the relevant scale) is much larger than the lattice spacing.
Then the angular field $\vp (x)$ of the simplest (rotationally symmetric) 
V or AV, with its core at $x=0$, obeys
\be  \label{nababs}
| \vec \nabla \vp (x)| = \frac{1}{r} \ , \quad r = |x| \ ,
\ee
with opposite gradient directions for a V or an AV, see Fig.\ 
\ref{vort}. In this continuum picture, the vorticity $v$ is given 
by a curl integral, anti-clockwise around the core,
\be  \label{v1}
v = \frac{1}{2\pi} \oint d\vec x \cdot \vec \nabla \vp (x) =
\frac{1}{2\pi} \int_{0}^{2\pi} r \, d\vp \ \Big( \pm \frac{1}{r} \Big)
= \pm 1 ~~{\rm for}~\left\{ \begin{array}{c} {\rm a~vortex} \\
{\rm an~anti\mbox{-}vortex.} \end{array} \right.
\ee
Regarding the energy, we note that the Hamilton function (\ref{Hami}) 
(at $\vec H = \vec 0$) can be considered as a kinetic term, made of
discrete derivatives,
\be  \label{Hamitrafo}
 J \sum_{\mu =1}^{2} ( 1 - \vec e_{x} \cdot \vec e_{x+ \hat \mu} )
\simeq \frac{J}{2} \sum_{\mu =1}^{2} \Delta \vp_{x, x+\hat \mu}^{2} 
\simeq \frac{J}{2} \, \vec \nabla \vp (x) \cdot \vec \nabla \vp (x) \ .
\ee
Here we switched from lattice to continuum notation, and 
we neglect $O(\Delta \vp_{x, x+\hat \mu}^{4})$.

If we insert relations (\ref{nababs}) and (\ref{Hamitrafo})
into the Hamilton function, we obtain an estimate for the energy 
requirement for inserting one V or AV into a smooth ``background'',
\be  \label{EV}
E_{\rm V} = \frac{J}{2} \int d^{2} x \ 
\vec \nabla \vp (x) \cdot \vec \nabla \vp (x)
\approx J \pi \int_{1}^{L} dr \ \frac{1}{r} = J \pi \ln L \ .
\ee
Note that (despite the continuum notation) $L$ expresses the system 
size in lattice units, so it is dimensionless (and taking its 
logarithm makes sense). The integral over the plane is a bit sloppy 
regarding the shape of the volume; it is approximated by 
a circle of radius $L$, except for a small inner disc with the 
radius of one lattice spacing (which we have set to 1). 
The latter matches the illustrations in Fig.\ \ref{vort}, and
such an UV cutoff is needed to obtain a finite result.

Even this simplified consideration captures relevant properties. 
The energy for a single V or AV is considerable:
it is enhanced $\propto \ln L$,
so it takes a high temperature to make such vortex 
excitations frequent. In the thermodynamic limit, $L \to \infty$,
they seem to be excluded, but we will see in Section 2.2 why
the topological defects are so important nevertheless.
{\em Vadim L.\ Berezinski\u{\i}}
(1935-80) explored these properties in 1970/1 \cite{VLB}. 
He was working in Moscow, where he pioneered the vortex 
picture \cite{obi}, which was later an inspiration
in the search for a confinement mechanism in QCD \cite{Poly92}.

\subsection{Refined picture}

\begin{figure}[h!]
\begin{center}
\includegraphics[width=0.45\textwidth,angle=0]{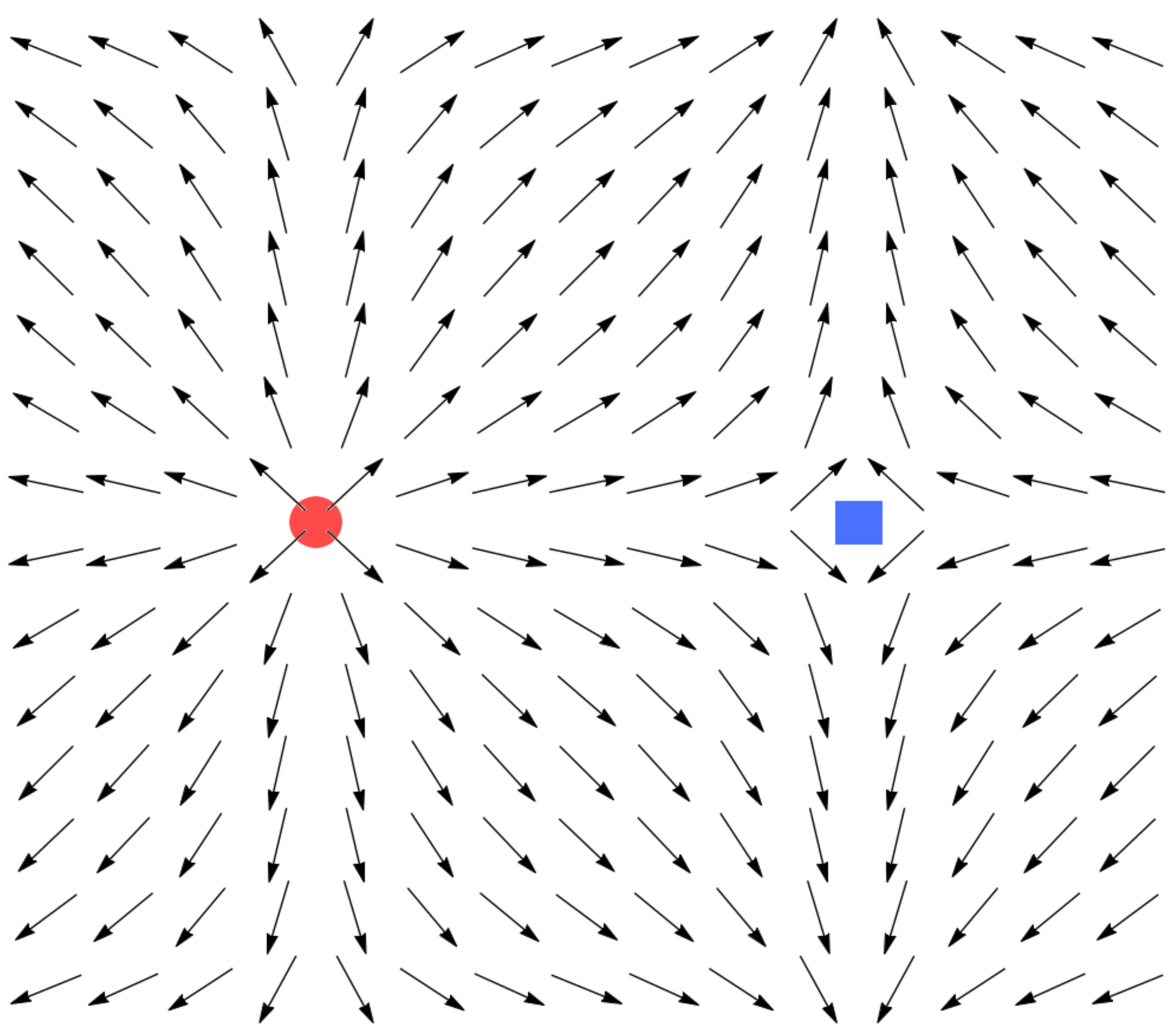}
\caption{Profile of a configuration with a V--AV pair, with
zero total vorticity: the V (AV) core is indicated by a red dot 
(blue square). Its energy is estimated in eq.\ (\ref{EVAV}).}
\label{VAVpair}
\end{center}
\vspace*{-5mm}
\end{figure}

The picture of Section 2.1 can be criticized for assuming
either a single V or a single AV in the entire configuration,
although we stressed before that their number must be equal
(with periodic boundaries). So the minimal excitation of
topological defects
leads to one V plus one AV, as illustrated in Fig.\ \ref{VAVpair},
and the above calculation has to be revised. In fact, the result 
is not $E_{\rm V,AV}^{\rm isolated} = 2E_{\rm V} = 2 \pi J \ln L$, 
but instead
\be  \label{EVAV}
E_{\rm V,AV} = 2 \pi J \ln r_{\rm V,AV} \ ,
\ee
where $r_{\rm V,AV}$ is the distance between the V and AV core.
This can be understood qualitatively: if the V--AV pair
is tightly bound, its long-distance impact cancels; far away, 
the configuration can be practically uniform, as in
the absence of any vortices. If we observe the system
with a low resolution (corresponding to a large $\xi$),
we might not see this pair at all.

Only pulling them far apart leads to {\em ``free''} V and AV, which
are visible to such an observer. When $r_{\rm V,AV}$ reaches the
magnitude of $L$, the energy requirement is of the 
order of $E_{\rm V,AV}^{\rm isolated}$. 

From eq.\ (\ref{EVAV}) we see that the trend towards minimal energy
implies an {\em attractive force} $\propto 1/r_{\rm V,AV}$
between the V and AV cores.
In $d=2$ this is a Coulomb force, so a few V and AV spread over the 
plane can be considered as a {\em Coulomb gas}. Its free energy $F$ 
consists of the total energy $E$, plus an entropy term (cf.\ Section 1).

In the period 1972-4, {\em John M.\ Kosterlitz} (born 1942 in Aberdeen), 
and {\em David J.\ Thouless} (born 1934 in Bearsden), both from Scotland, 
worked on this issue at the University of Birmingham. 
They concluded that the driving
force of the transition between the massive and the massless phase
is not exactly the density $\rho$ (referred to in Section 2.1), but 
the density of ``free vortices and anti-vortices'',  {\it i.e.}\ 
V or AV without any opposite partner nearby. So the phase transition 
is actually {\em driven by the (un-)binding of V--AV pairs} \cite{KT}.

To make this picture more explicit, we consider 
the free energy $F$, say in a sub-volume which
is large enough to accommodate one free V. It is convenient to
call its size $L$, and to recycle formula (\ref{EV}).
The entropy $S$ is the logarithm of the multiplicity of such 
configurations, here this is just the number of $L^{2}$ plaquettes 
where the vortex could be located. This yields
\be  \label{FlnL}
F = E_{\rm V} - T S = J \pi \ln L - T \ln L^2 
= (J \pi - 2T) \, \ln L \ ,
\ee
and the phase of the system depends on the question which of these two 
terms dominates. 

\begin{itemize}

\item At low $T$ there are hardly any free V or AV (they are suppressed 
when $L$ becomes large), though there might be some tight V--AV pairs. 

\item At high $T$ these pairs unbind: due to the dominance of the second term,
a large size $L$ makes it easy to spread free V and AV all
over the system. 

\end{itemize}

In this setting, eq.\ (\ref{FlnL}) suggests that the critical temperature, 
where the transition happens, amounts to $T_{\rm c} = J \pi /2$ 
\cite{KT}.

\subsection{Critical behavior}

Kosterlitz and Thouless predicted a type of phase transition, which 
had been unknown before the 1970s. The correlation length diverges
when $T$ decreases down to $T_{\rm c}$, as in the well-known phase 
transitions of second order, but in contrast to them $\xi = \infty$
persists at $T < T_{\rm c}$, and {\em no symmetry breaking is involved.}
This means a step beyond Landau's Theory, which successfully describes 
second order phase transitions with the concept of spontaneous symmetry 
breaking. In low dimensions ($d\leq 2$), however, thermal
fluctuations are powerful enough to prevent spontaneous ordering,
like a magnetization $M>0$. This has been demonstrated
generally by the Mermin-Wagner Theorem \cite{MW}, 
and specifically for the 2d O($N$) models in Ref.\ \cite{Wegner}.
The characteristics of the BKT transition were also confirmed experimentally,
in particular in thin films of superfluid $^{4}$He \cite{superfluid} and of
superconductors \cite{supercon}. 

With respect to the critical exponents, this transition was discussed
comprehensively by Kosterlitz in 1974 \cite{Kos74}, based on
Renormalization Group techniques.
He pointed out that this is a phase transition of infinite order, 
an {\em essential phase transition}. 
The correlation length $\xi$ is not described by a power divergence
as in relation (\ref{criteta}), but by an essential singularity,
\be \label{xiess}
\xi \propto \exp \Big( \frac{\rm const.}
{(T - T_{\rm c})^{\nu_{\rm e}}} \Big) \ ,
\quad T \gtapprox T_{\rm c} \ .
\ee
Thus one defines a critical exponent $\nu_{\rm e}$ for
the exponential growth of $\xi$; Kosterlitz derived its value 
$\nu_{\rm e} = 1/2$.

Since there is no symmetry breaking going on in the BKT transitions,
we cannot address the critical exponent $\beta$, and the
susceptibility $\chi_{\rm m}$ does not follow relation (\ref{gamma})
either. The critical exponents of
Section 1.1 all refer to infinite volume, but in the 2d XY model at
$V = L \times L \to \infty$, $\chi_{\rm m}$ diverges
throughout the massless phase. Kosterlitz predicted how it 
diverges as a function of $L$ (the scale which is left) \cite{Kos74},
\be  \label{chicrit}
\chi_{\rm m} \propto L^{2-\eta_{\rm e}} (\ln L)^{-2 r_{\rm e}} \ ,
\quad \eta_{\rm e} = 1/4 \ , \ r_{\rm e} = - 1/16 \ .
\ee
This prediction is hard to verify numerically: studying the 
logarithmic term (and further sub-leading logarithms)
requires huge volumes. A particularly extensive investigation
with the standard Hamilton function (\ref{Hami}) in Ref.\ \cite{Has} 
(see also Ref.\ \cite{Jap}) is based on 
simulations up to size $L=2048$, and the outcome is
consistent with the predicted exponents $\eta_{\rm e}$ and $r_{\rm e}$,
though $r_{\rm e}$ comes with a large error.

At this point we mention an alternative and entirely
different Hamilton function for the O($N$) spin models. 
Unlike the term (\ref{Hami}), it does not include any
(discrete) derivative term, but just a cutoff $\delta$ 
for the angular difference between any two nearest neighbor spins,
\be  \label{Hcon}
{\cal H} [\vec e \, ] = \left\{ \begin{array}{ccc}
0 && {\rm if} ~~ |\Delta \vp_{x,x+\hat \mu} | < \delta ~~ \forall x, \mu \\
\infty && {\rm otherwise.} \end{array} \right.
\ee
Such a {\em constraint Hamilton function} is 
{\em topologically invariant,} which means that
most small modifications of a configuration leave the energy
exactly constant. This is highly unusual:
part of the configurations are excluded (those that violate the
constraint), while all others have energy 0. 
Still, it has the same symmetries as the standard action, and 
it belongs to the same universality class \cite{topact,O2top,BKT}. 

There is no temperature in this formulation, but the constraint
angle $\delta$ plays a r\^{o}le, which bears some analogy.
In fact, there is a critical $\delta_{\rm c}$, and the system is
in its massive (massless) phase for $\delta > \delta_{\rm c}$ 
($\delta < \delta_{\rm c}$). When $\delta$ approaches its critical 
value within the massive phase, the correlation length exhibits
an exponential divergence as in relation (\ref{xiess}) \cite{O2top},
\be  \label{xidiv}
\xi \propto \exp \Big( \frac{\rm const.}
{(\delta - \delta_{\rm c})^{\nu_{\rm e}}} \Big) \ ,
\quad \delta \gtapprox \delta_{\rm c} \ .
\ee
\begin{figure}[h!]
\centering
\includegraphics[width=0.55\textwidth,angle=0]{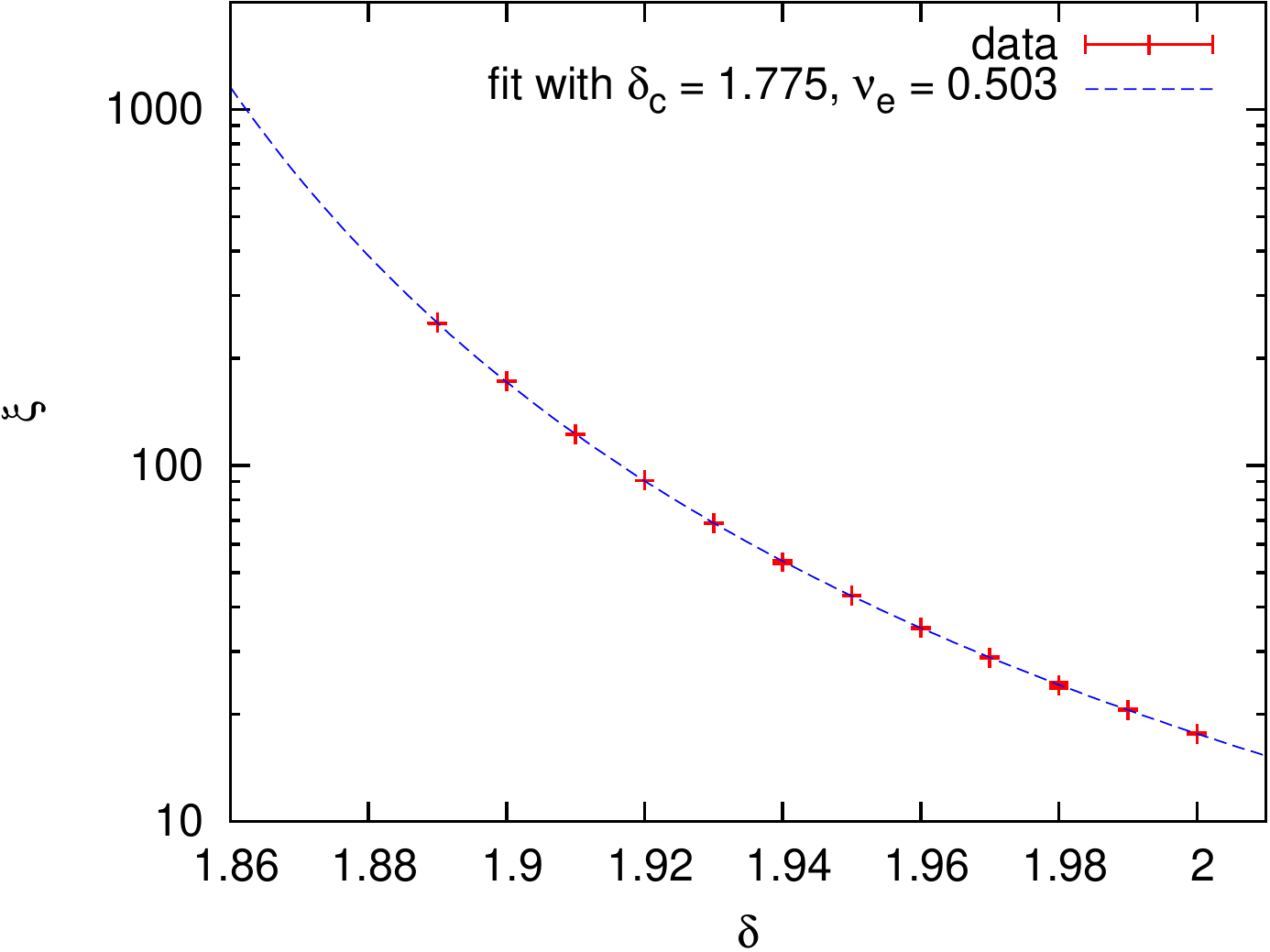}
\vspace*{-2mm}
\caption{The exponential divergence of the correlation length 
$\xi$, as $\delta$ decreases towards $\delta_{\rm c}\simeq 1.775$.
A fit to relation (\ref{xidiv}) confirms Kosterlitz' prediction
for $\nu_{\rm e}$.}
\label{xidelta}
\end{figure}
This observation singles out the critical constraint angle $\delta_{\rm c} 
= 1.775(1)$. Fig.\ \ref{xidelta} shows this divergence as 
$\delta > \delta_{\rm c}$
decreases, and the fit yields $\nu_{\rm e} = 0.503(7)$, accurately
confirming Kosterlitz' prediction.

Regarding the limit within the massless phase, the divergence 
of the susceptibility $\chi_{\rm m}$ is consistent with the relation
(\ref{chicrit}), and $\eta_{\rm e}$ is confirmed to two digits 
\cite{O2top}, whereas the value for $r_{\rm e}$ is plagued by large 
uncertainties, as in Ref.\ \cite{Has}.\\

Another prediction for the BKT transition in the 2d XY model refers to
the {\em helicity modulus} (or {\em spin stiffness}).  
In its dimensionless form, it is defined as
\be
\Upsilon = \frac{1}{T} \, \frac{\partial^{2}}{\partial \alpha^{2}} 
F |_{\alpha = 0} \ ,
\ee
where $\alpha$ is a twist angle in the boundary conditions. The free
energy $F$ is minimal at $\alpha =0$ (periodic boundaries), and
$\Upsilon$ is the curvature in this minimum.

\begin{figure}[h!]
\begin{center}
\includegraphics[width=0.48\textwidth,angle=0]{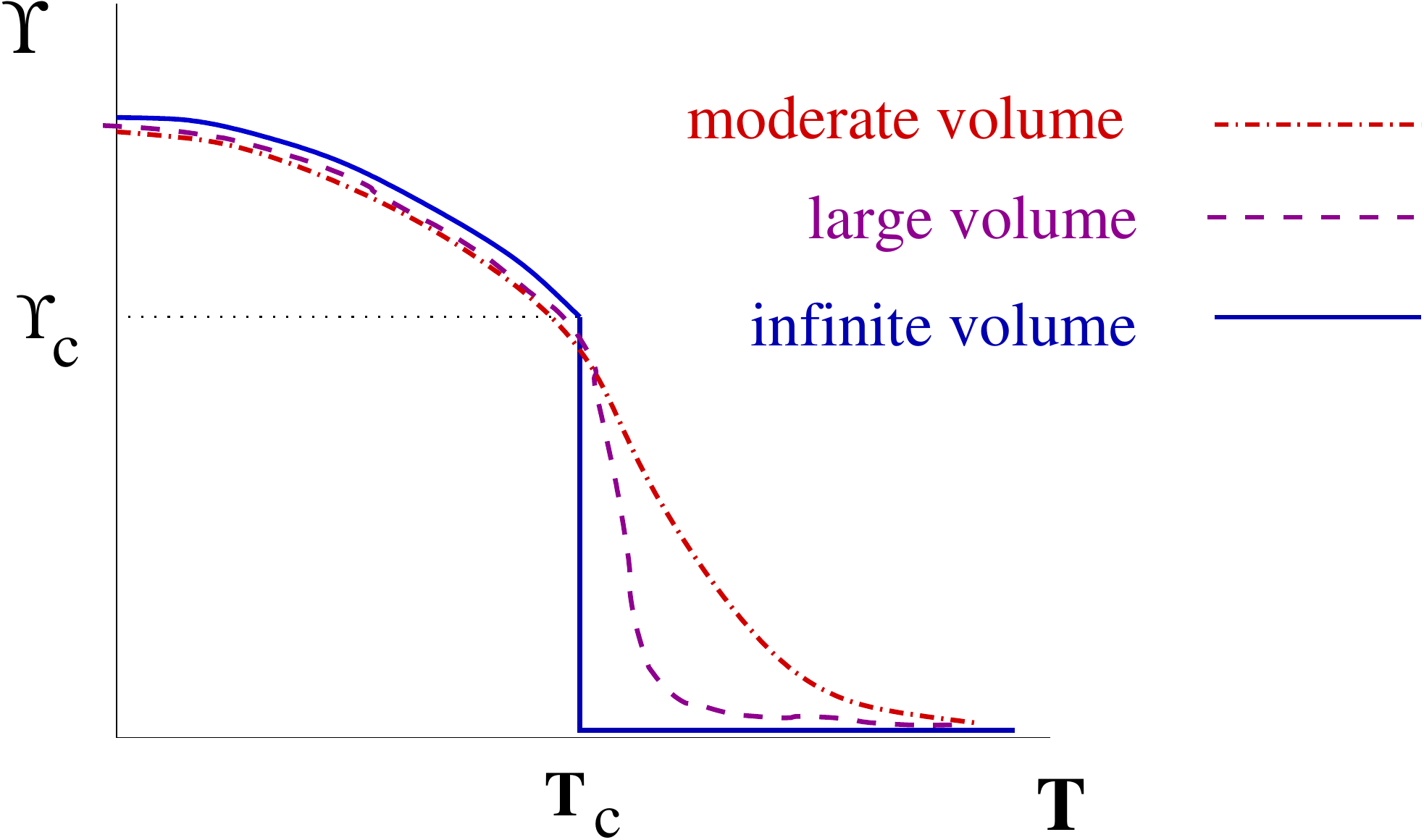}
\hspace*{3mm}
\includegraphics[width=0.47\textwidth,angle=0]{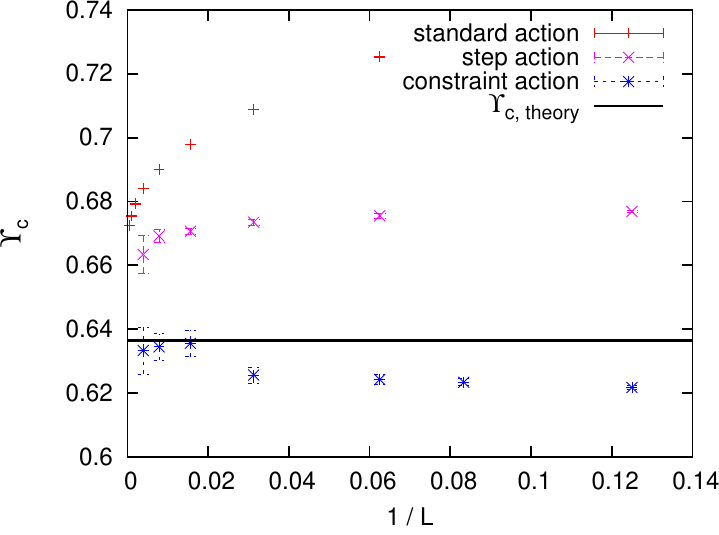}
\caption{A qualitative picture of the helicity modulus $\Upsilon$
depending on the temperature (left), and an overview over numerical 
results for its helicity jump at the critical point, $\Upsilon_{\rm c}$
(right).}
\label{Yjump}
\end{center}
\vspace*{-5mm}
\end{figure}

The qualitative picture is illustrated in Fig.\ \ref{Yjump} (left): 
in the large volume limit, one expects $\Upsilon$ to perform
a universal jump at $T_{\rm c}$ \cite{JKKN}. Soon after 
the BKT transition had been put forward, the height of this jump was
predicted as $2/\pi$ \cite{NelKos}. 
Later a small correction was subtracted \cite{ProSvi} to arrive
at the theoretical value
\be  \label{Upstheo}
\Upsilon_{\rm c,theory} = \frac{2}{\pi} \Big( 1 - 16 e^{-4\pi} \Big)
\simeq 0.6365 \ .
\ee
Regarding the constraint Hamilton function, we can interpret 
$\frac{1}{Z} \exp(-F(\alpha)/T)$ generally as the probability 
for a (dynamical) twist angle $\alpha$, so the helicity modulus 
can be studied without needing the concept of temperature.

Fig.\ \ref{Yjump} (right) summarizes simulation results for $\Upsilon_{\rm c}$ 
obtained with various lattice Hamilton functions. The standard formulation
(\ref{Hami}) is very tedious in this regard: even simulations at $L=2048$
yielded $\Upsilon_{\rm c} = 0.67826(7)$ \cite{Has}, which is far too high.
Somewhat more successful was the use of a ``step Hamilton function'', which is also 
topologically invariant: when $|\Delta \vp_{x,x+\hat \mu}|$ exceeds $\delta = \pi /2$, 
the energy contributions of this pair of neighboring spins jumps from zero
to some finite value, which is varied (instead of varying $\delta$).
Here $L=256$ led to  $\Upsilon_{\rm c} = 0.6634(6)$ \cite{OlHol}, but it still
took faith to accept the compatibility of the large-$L$ extrapolation
with the theoretical value in eq.\ (\ref{Upstheo}).

This compatibility was finally demonstrated beyond doubt with the
constraint Hamilton function (\ref{Hcon}). As a function of $\delta$
(replacing $T$), $\Upsilon$ behaves exactly as depicted in 
Fig.\ \ref{Yjump} (left): a jump is observed around $\delta_{\rm c}$, 
and it becomes more marked as the volume increases.
At $\delta_{\rm c}$ the value $\Upsilon_{\rm c} = 0.636(4)$ was 
measured already at $L=64$, and larger volumes confirmed the agreement
with eq.\ (\ref{Upstheo}) \cite{BKT}. This is one of the clearest pieces 
of numerical evidence that the BKT transition does occur, and that the 
corresponding quantitative predictions are valid.\footnote{Alternative
numerical evidence is obtained from the Step Scaling Function \cite{SSF,O2top}. 
It expresses the change of the ratio $L/\xi$ when the size $L$ is altered, at 
fixed (nearly critical) $T \gtapprox T_{\rm c}$ or $\delta\gtapprox \delta_{\rm c}$.}

The height of this jump, $\Upsilon_{\rm c}$, is related to the essential 
critical exponent $\eta_{\rm e}$. It can be translated into the jump 
of the {\em superfluid density} \cite{NelKos}, which has been observed 
in films of\, $^{4}$He \cite{superfluid}, and recently also in an
optically trapped 2d Bose gas \cite{Kor}.\\

All this seems nicely consistent, but in some sense it is {\em puzzling:}
in Section 2.2 we reviewed the consideration of energy vs.\
entropy in the vortex picture, which predicts the BKT transition.
This picture is standard, and it has been 
brought into further prominence by the Nobel Prize Committee.
However, in the formulation with the constraint Hamilton function 
the energy cost for any V or AV is {\em zero,} 
but still the BKT transition is beautifully observed \cite{O2top,BKT}. 
Is this a contradiction?
\begin{figure}[h!]
\begin{center}
\includegraphics[width=0.325\textwidth,angle=0]{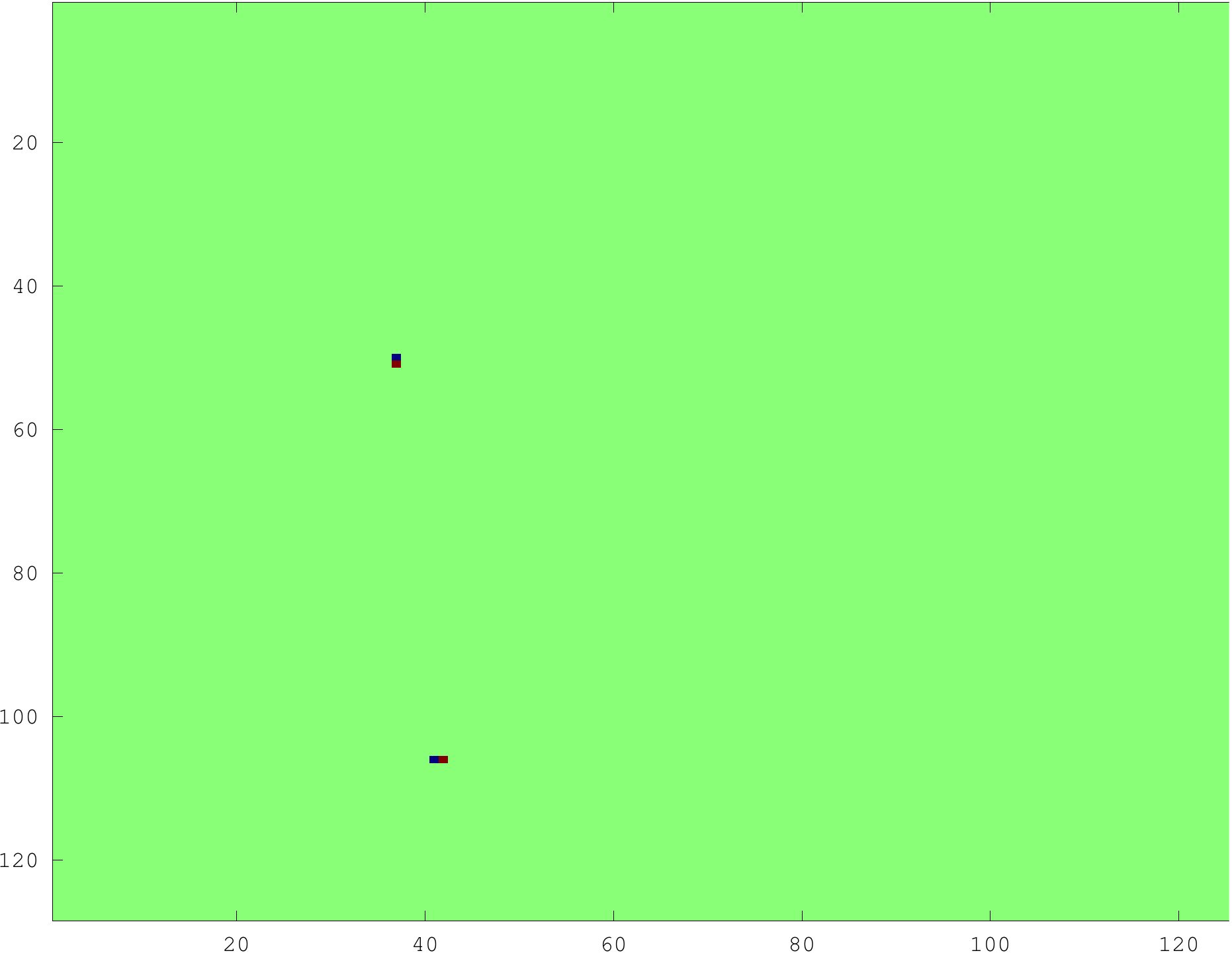}
\includegraphics[width=0.325\textwidth,angle=0]{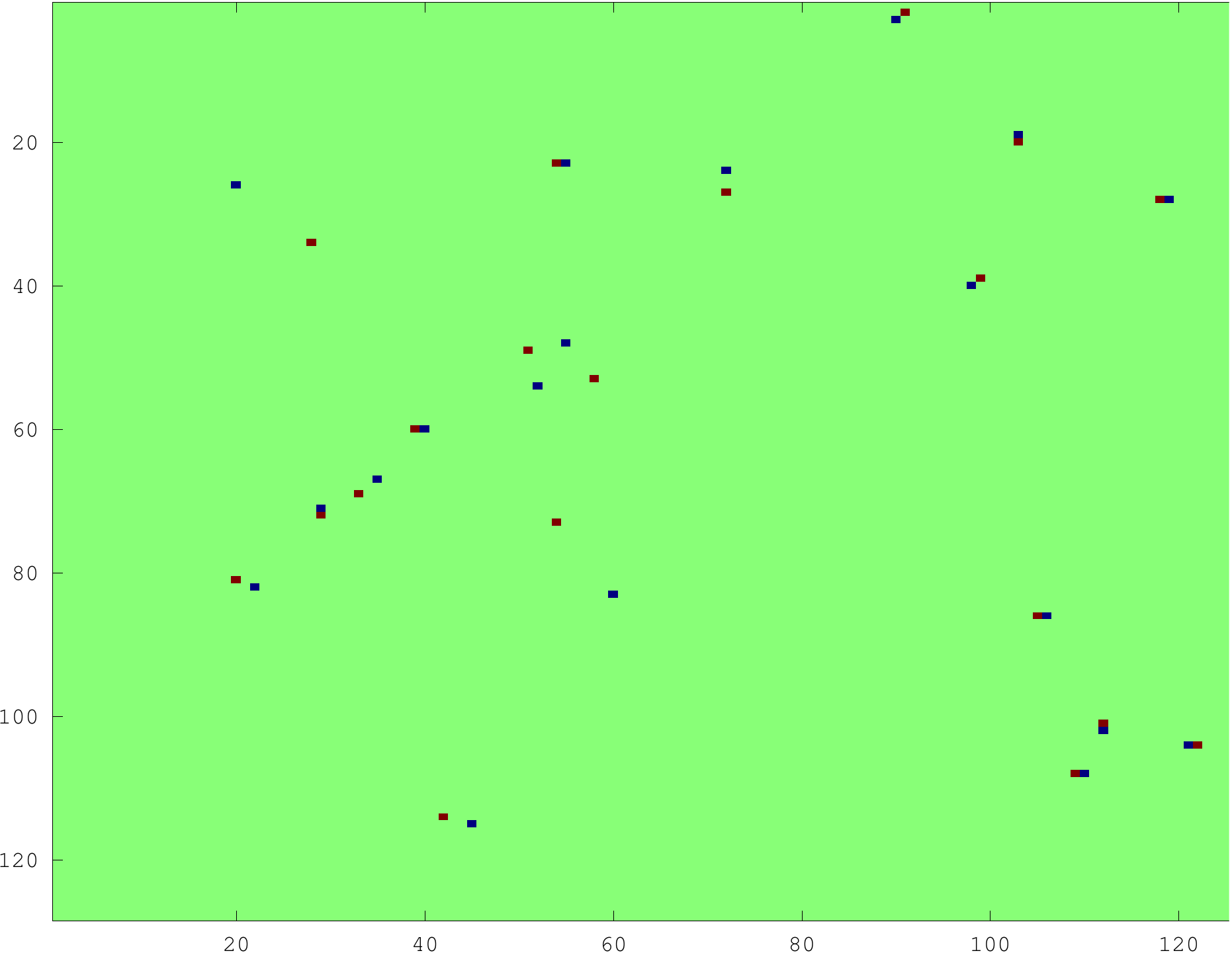}
\includegraphics[width=0.325\textwidth,angle=0]{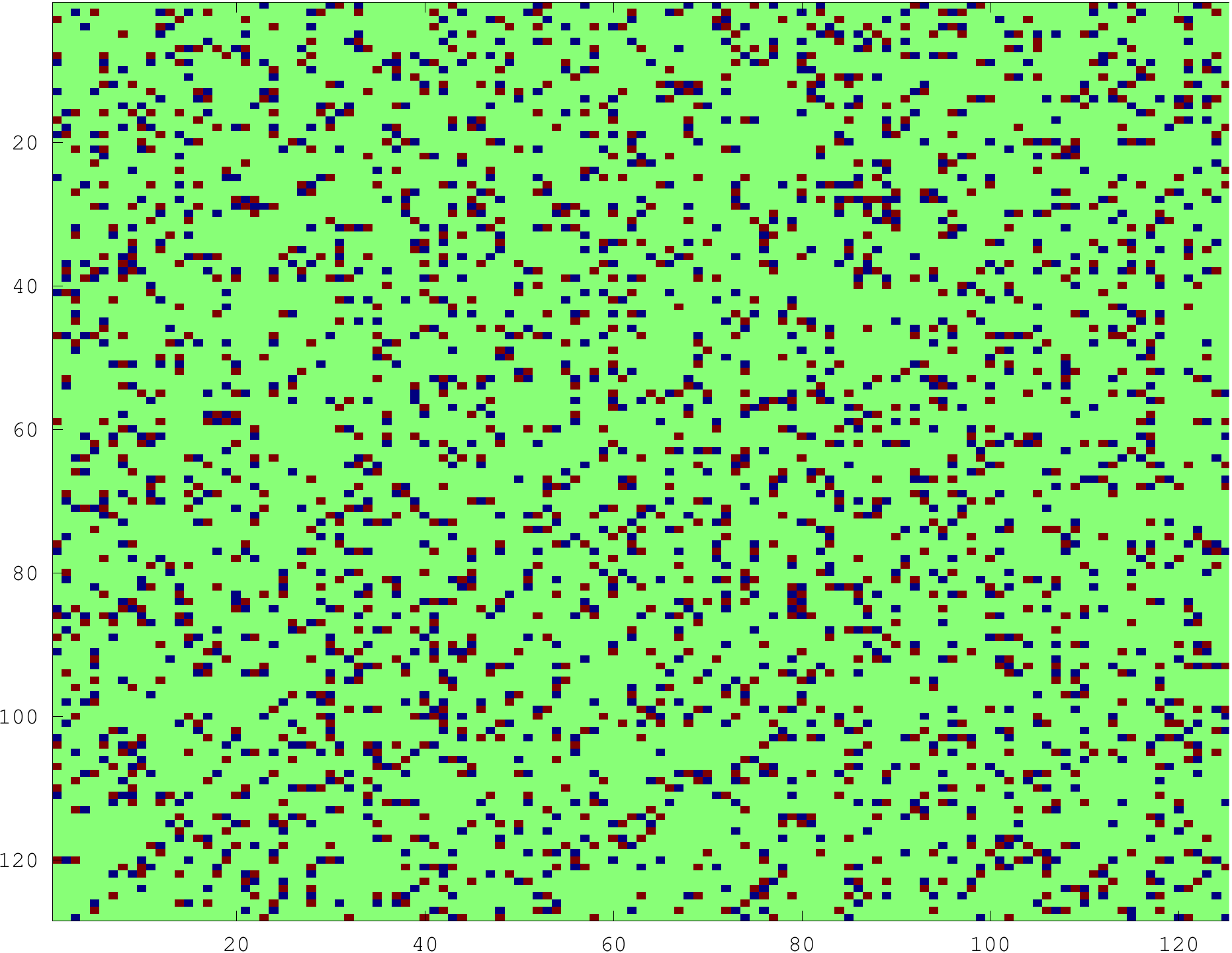}
\caption{Maps of typical configurations of the XY model on a 
$128 \times 128$ lattice, with the constraint Hamilton function 
(\ref{Hcon}) and $\delta =1.8$, $1.9$ and $2.5$ (from left to right). 
The presence of a V (AV) is indicated by a red (blue) plaquette.
As long as there are only few V (for $\delta \leq 1.9$), 
the effect of V--AV pair formation is evident.}
\label{Vmap}
\end{center}
\vspace*{-5mm}
\end{figure}

More to the point, we focus on the question: is the V--AV (un-)binding
mechanism still at work, even when free V and AV do not cost any
energy?

A first hint is given by Fig.\ \ref{Vmap}, 
which shows ``maps'' of the V and AV found in
typical configurations at $\delta = 1.8, \ 1.9$ and $2.5$.
For small $\delta$, when only few V and AV show up, the trend to a
V--AV pair formation is obvious. At $\delta =2.5$ there are numerous
topological defects, and it cannot be seen by eye whether or not
such a trend persists.

\begin{figure}[h!]
\begin{center}
\includegraphics[width=0.49\textwidth,angle=0]{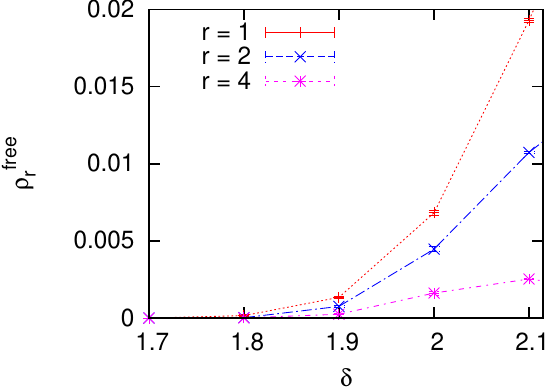}
\hspace*{3mm}
\includegraphics[width=0.43\textwidth,angle=0]{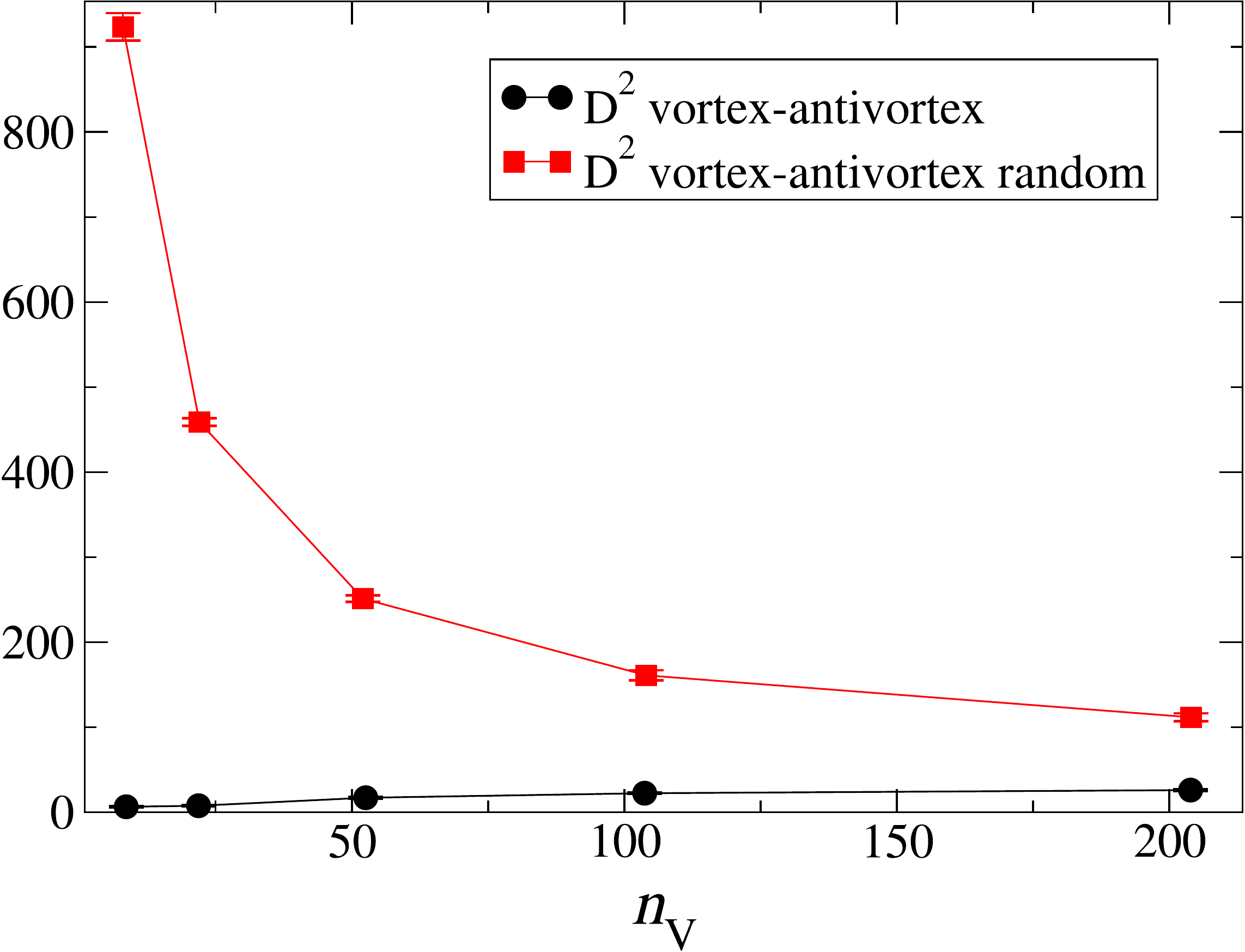}
\vspace*{-3mm}
\caption{Left: The density of ``free vortices'', $\rho_{r}^{\rm free}$,
{\it i.e.}\ of V or AV without an opposite partner within distance
$r=1,\, 2$ or $4$. We see an onset at $\delta \protect\gtapprox 
\delta_{\rm c}$, so $\rho_{r}^{\rm free}$ is similar to an 
(inverse) order parameter.
Right: The mean distance squared between V--AV pairs, $D^{2}$,
for optimal pairing (black line). For small $n_{\rm V}$ (few vortices), 
$D^{2}$ is {\em much} shorter than the corresponding term for
random distributed V and AV (red line). Around 
$n_{\rm V} \protect\gtrsim 50$ (typical for $\delta \approx 1.9$) 
this striking discrepancy fades away. This confirms the V--AV 
(un-)binding mechanism in the BKT transition.}
\label{Vsep}
\end{center}
\vspace*{-5mm}
\end{figure}

In any case, Fig.\ \ref{Vmap} only shows specific configurations, but 
a conclusive answer requires a statistical analysis. Fig.\ 
\ref{Vsep} (left) shows the average density $\rho_{r}^{\rm free}$ of ``free V'' 
plus ``free AV'', defined by the property that there is no opposite
partner within distance $r$, with $r = 1,\, 2$ and $4$ (at $L=128$).
We see an onset around $\delta \simeq 1.8$, and a sharp increase 
as $\delta$ exceeds $1.9$. Hence $\rho_{r}^{\rm free}$ behaves indeed 
like an (inverse) ``order parameter'' for the BKT transition\footnote{The 
finite volume shifts the apparent critical angle somewhat up.}
(although, strictly speaking, there is no ordering).

Fig.\ \ref{Vsep} (left) shows the mean distance squared between 
nearby V and AV cores, $d^{2}_{\rm V,AV}$, in configurations with 
$n_{\rm V}$ vortices (and $n_{\rm V}$ anti-vortices), also at $L=128$,
\be
D^{2} = \frac{1}{n_{\rm V}} \sum_{i=1}^{n_{\rm V}} d^{2}_{{\rm V,AV},i} \ .
\ee
The V and AV pairs are formed such that $D^{2}$ is minimal.
This is compared to $D^{2}$ for artificial configurations, where 
the same number of V and AV are random distributed over the volume.
For small $n_{\rm V}$ --- which corresponds to small $\delta$ ---
we see a striking difference for the configurations which are 
generated by simulating the model. 
This is clear evidence for a V--AV pair formation.
This effect fades away at larger $\delta$, when $n_{\rm V}$
increases ($\delta = 1.9$ corresponds to about $n_{\rm V}=50$).

We conclude that the V--AV (un-)binding mechanism is at work, which
confirms once more the elegant picture by Kosterlitz and Thouless 
for the BKT phase transition.
This observation holds {\em even} when topological defects do not cost 
any energy; then it is a pure entropy effect. Therefore the standard 
argument for this picture  --- outlined in Section 2.2 ---
should be extended.

\begin{figure}[h!]
\vspace*{1mm}
\begin{center}
\includegraphics[width=0.22\textwidth,angle=0]{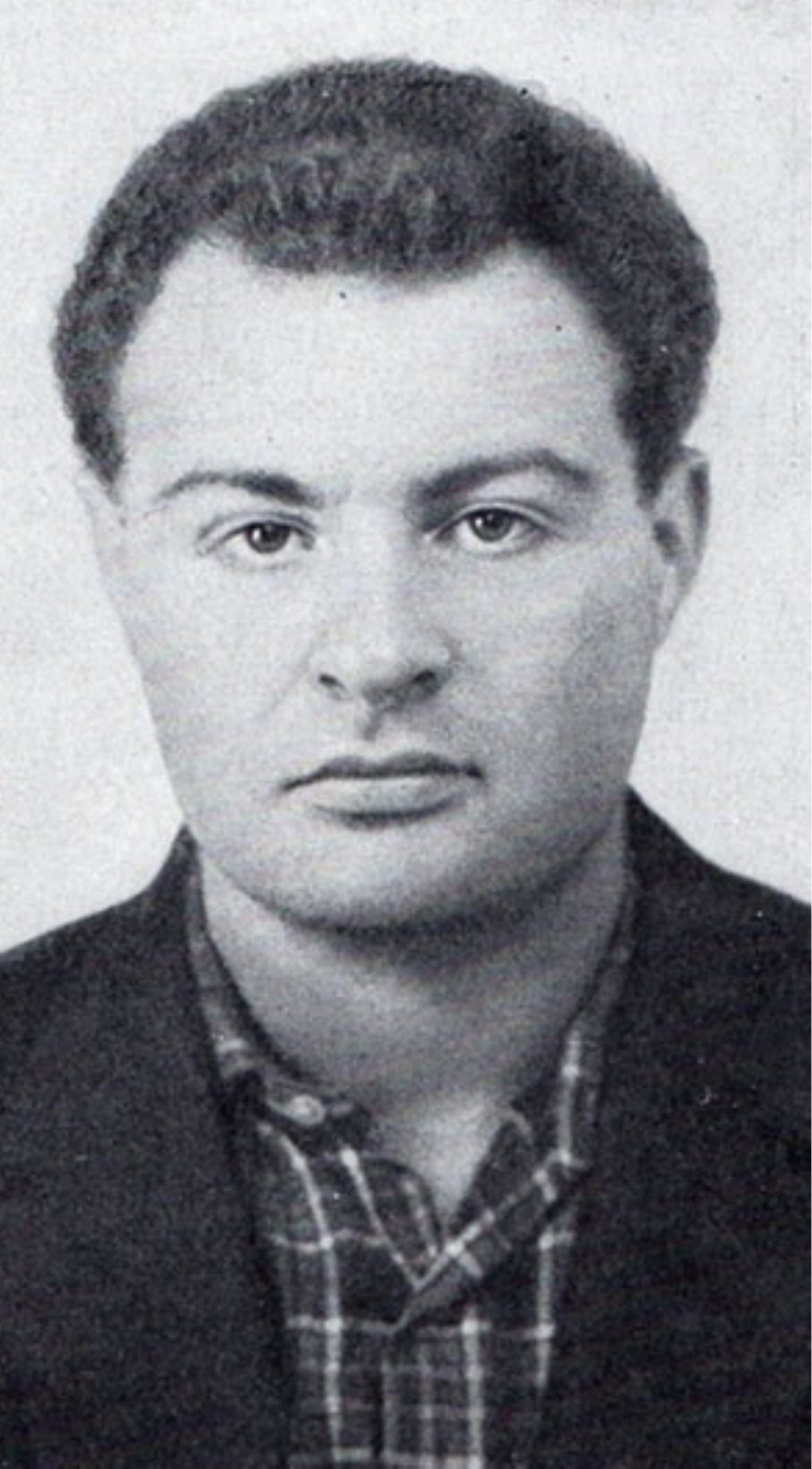}
\hspace*{3mm}
\includegraphics[width=0.43\textwidth,angle=0]{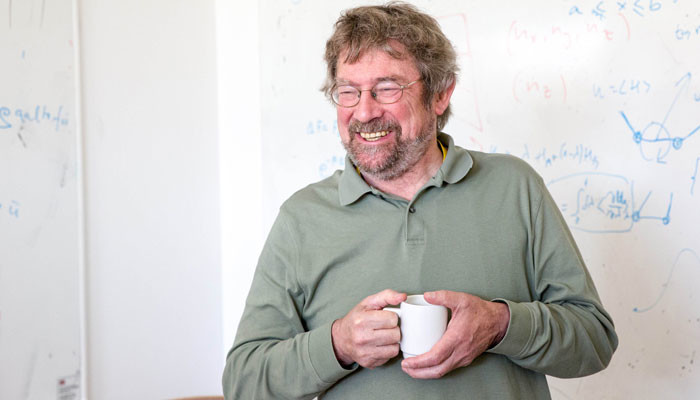}
\hspace*{3mm}
\includegraphics[width=0.25\textwidth,angle=0]{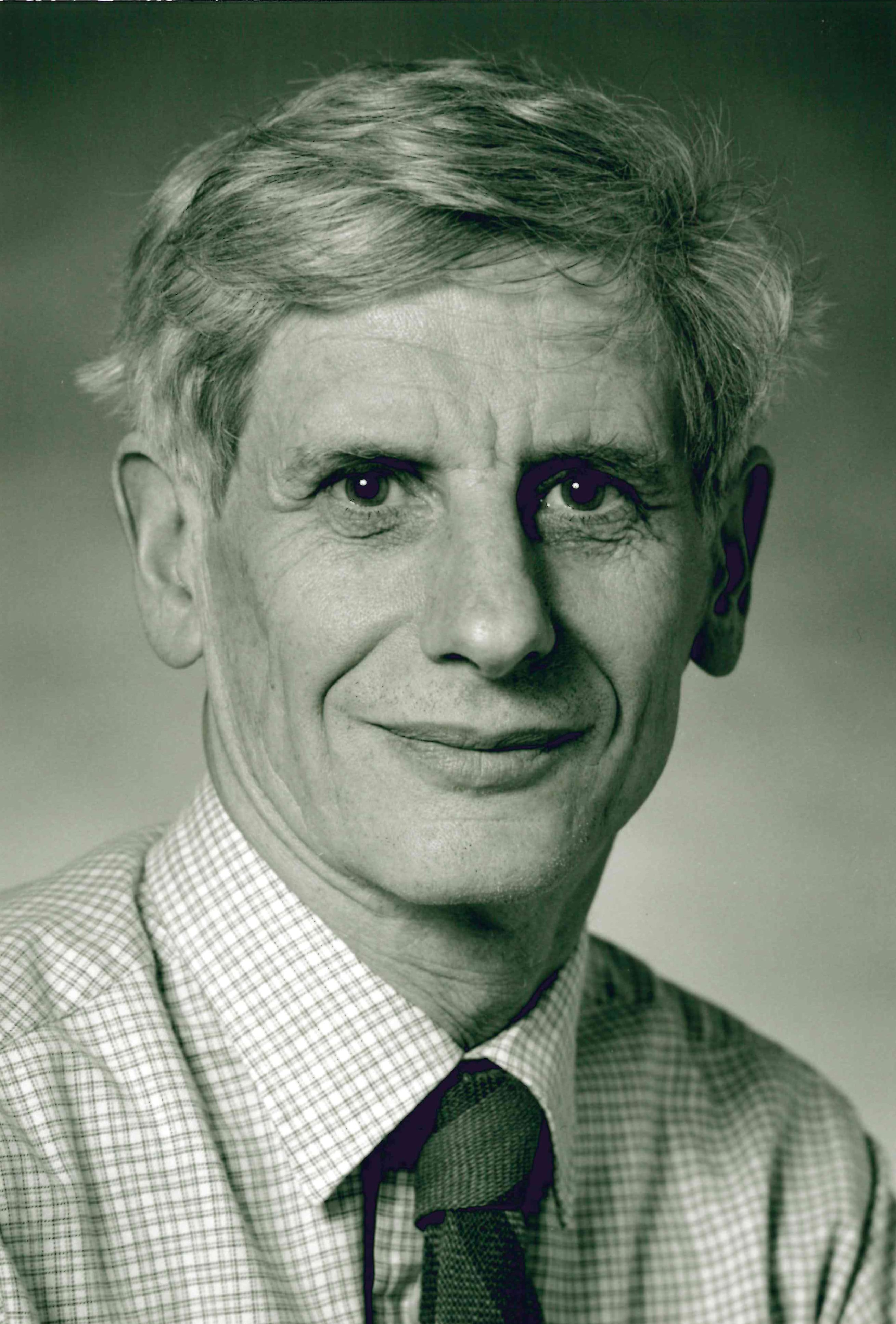}
\vspace*{-3mm}
\end{center}
\caption{From left to right:
{\bf Vadim L'vovich Berezinski\u{\i}} (1935-1980) was born in Kiev (USSR)
and graduated 1959 at Moscow State  University. 
After working at the Textile Institute and 
the Research Institute for Heat Instrumentation, he joined 1977 the Landau
Institute of Theoretical Physics in Moscow.
{\bf John Michael Kosterlitz} was born 1942 in Aberdeen (Scotland),
studied at Cambridge University, and graduated 1969 in Oxford.  
In 1974 he become Lecturer at Birmingham
University, and in 1982 Professor at Brown University in
Rhode Island, USA.
{\bf David James Thouless} was born 1934 in Bearsden (Scotland).
He studied at Cambridge University as well, and graduated 1958 
at Cornell University, his Ph.D. advisor was Hans Bethe.
He worked in Birmingham with  Rudolf Peierls,
and later with John Kosterlitz. In 1980 he became Professor
at the University of Washington in Seattle.}
\label{bio1}
\vspace*{-5mm}
\end{figure}

\section{Haldane Conjecture}

We now proceed to {\em quantum spin models,} leaving behind the 
classical spins (albeit they will be back in Section 3.1).
Now the components of a spin vector are Hermitian {\em operators,}
for spin $1/2$ they can be represented by the Pauli matrices.
For any spin, $s = 1/2,\, 1, \, 3/2, 2, 5/2 \dots$
(in natural units, $\hbar =1$), we write them as $\hat S^{a}_{x}$, 
where $x$ is still a lattice site. These components
obey the familiar relations
\be
[ \hat S^{a}_{x}, \hat S^{b}_{y} ] = \ri \, \delta_{xy} \, 
\epsilon^{abc} \, \hat S^{c}_{x} \ , \quad
\sum_{a=1}^{3} \hat S^{a}_{x} \, \hat S^{a}_{x} = s(s+1) \ ,
\ee
where $\epsilon$ is the Levi-Civita symbol.
If we compare these terms at large $s$, we see that the 
commutator is suppressed as $O(s) \ll O(s^{2})$, and the
spin appears nearly classical.

For arbitrary spin we assemble the {\em Hamilton operator} $\hat H$,
and write down the partition function,
\be  \label{qHami}
\hat H = - J \sum_{\la xy \ra, a} \hat S^{a}_{x} \, \hat 
S^{a}_{y} \ , 
\quad Z = {\rm Tr} \ e^{- \hat H /T} \ .
\ee
It is analogous to the Hamilton function (\ref{Hami}) and 
partition function (\ref{prob}), now with quantum spins.
We recognize a global SU(2) symmetry, which transforms 
each component $\hat S^{a}_{x}$.

In addition, the current framework differs 
from the previous sections in the following points: 
\begin{itemize}

\item We focus on {\em spin chains,} {\it i.e.}\ dimension 
$d=1$, so now the sites are located on a line.

\item We consider {\em anti-ferromagnets,} with $J<0$,
cf.\ footnote \ref{affoot}.

\item We skip the external magnetic field.

\item We drop the additive constant $JVd$ (``cosmological constant'')
of the Hamilton function  (\ref{Hami}).
This change is irrelevant --- what matters are solely energy 
{\em differences.}

\end{itemize}

For commutative spin components it would be trivial to write down 
a ground state of such an anti-ferromagnetic spin chain: it
consists of spins of opposite orientations, in alternating order
(say $|s, \, -s, \, s, \, -s, \, s, \, -s \dots \ra \,$),
known as a N\'{e}el state.
However, this is not an eigenstate of $\hat H$.
Quantum spins are far more complicated, and identifying a 
ground state is a formidable task, even in $d=1$.

The investigation of these systems has a history of almost 100 years.
The ongoing interest has been fueled by the fact that quantum spin 
chains exist experimentally; we will give examples below.
A breakthrough was achieved by Hans Bethe in 1931, who
constructed the ground state for spin $s=1/2$ \cite{Bethe}.
 
Of course also excited states are
of interest, and in particular the question whether or not
there is a finite energy gap $\Delta_{s} = E_{1}-E_{0}$. We
repeat that a finite gap corresponds to a massive phase, 
with a correlation length $\xi = 1/\Delta_{s}$.

In the 1950s and 1960s such systems were studied mostly with
``spin wave theory'', an approach which was fashion at that time.
It predicts a ``quasi long-range order'' (without Nambu-Goldstone
bosons), which means a power decay of the correlation function, 
{\it i.e.}\ the massless case with $\xi = \infty$. This was 
elaborated mostly in higher dimensions, $d\geq 2$, doubts 
remained about the spin chain.

For $d=1$, the expected zero gap for $s=1/2$ was proved in 1961
by the Lieb-Schultz-Mattis Theorem \cite{LSM}. This consolidated
the paradigm that anti-ferromagnetic quantum spin chains are 
always gapless, for any spin 
$s = 1/2,\, 1,\, 3/2 \dots$ .

Therefore it came as a great surprise when
{\em F.\ Duncan M.\ Haldane} (born 1951 in London)
contradicted in 1983 \cite{Haldane1,Haldane2}. According to the 
{\em Haldane Conjecture,} the paradigm was correct only for 
the half-integer spins, 
but not for $s \in \NN \, $. He conjectured
\bea
s = 1/2, \ 3/2,\ 5/2 \dots & \mbox{(half-integer)} 
& \Delta_{s}=0 \quad \mbox{gapless} \nn \\
s = 1, \ 2,\ 3 \dots & \mbox{(integer)} 
& \Delta_{s}>0 \quad \mbox{finite~gap.} \label{HC}
\eea
Haldane gave topological arguments,
which were considered as somewhat cryptic, hence they were
initially met with skepticism. We refrain from an attempt 
to review them, here we refer to Ref.\ \cite{Fradkin}.

The zero gap for all half-integer spins was rigorously proved three 
years later \cite{AffLieb}, extending the  Lieb-Schultz-Mattis Theorem.

The surprising part of this conjecture, which refers to integer 
spins, was soon supported by numerical studies for $s=1$ 
\cite{BotJul}. Later the existence of a gap $\Delta_{1}>0$ was 
proved in a related system, where the Hamiltonian (\ref{qHami}) 
(Heisenberg Hamiltonian) was extended by the bi-quadratic term 
$-(J/3) \sum_{\la xy \ra } \, (\sum_{a} \hat S^{a}_{x} \hat S^{a}_{y})^{2}$ 
\cite{AKHT}. For the standard Hamiltonian (\ref{qHami}), the 
value of $\Delta_{1}$ was established numerically to 
high precision since the early 1990s \cite{White}. A study based on 
the diagonalization of an $L=22$ spin chain, and a large $L$ 
extrapolation, obtained $\Delta_{1} = 0.41049(2) \, J$ \cite{Gol}.

This is in agreement with {\em experimental} studies. 
In particular, the material Cs\,Ni\,Cl$_{3}$ contains 
quasi-1d anti-ferromagnetic $s=1$ spin chains.
The scattering of polarized neutrons
leads to a multi-peak structure, from which the value 
$\Delta_{1} \simeq 0.4 \ J$ could be extracted \cite{ex1}.
Similar observations were made with 
Ni(C$_{2}$H$_{8}$N$_{2}$)$_{2}$NO$_{2}$ClO$_{4}$ \cite{ex2},
but no gap was found in materials with $s=1/2$
spin chains \cite{spinhalf}. 

For higher $s \in \NN$, it is difficult to observe such a gap:
it has a conjectured extent
$\Delta_{s} \sim \exp(-\pi s)$ \cite{meronpic},
so it becomes tiny for increasing $s$. The case $s=2$ is still 
tractable numerically: a study up to $L=350$ 
arrived at $\Delta_{2} = 0.085(5) \, J$ \cite{Scholl}.\\
 
{\em In summary, the Haldane Conjecture (\ref{HC}) has been proved 
rigorously for all half-integer spins. For the integer spins there
are theoretical conjectures. In particular for  $\Delta_{1}$
they are supported by consistent numerical and experimental results.
In addition there is numerical evidence for $\Delta_{2} > 0$.}

\subsection{Mapping onto the 2d O(3) model}

A new perspective occurred by mapping such anti-ferromagnetic 
quantum spin chains onto the 2d O(3) model, or Heisenberg model. 
The latter emerged as a low energy effective theory, 
which was constructed by a large-$s$ expansion, and its validity
was conjectured for all $s$ \cite{Haldane2,Affmap} (for a review,
see Ref.\ \cite{Affrev}). 

Thus we are back with a classical spin model of Section 1. 
We write its Hamilton function in continuum notation,\footnote{The 
term in eq.\ (\ref{O3Hami}) is usually interpreted as an Euclidean 
action. Here it is embedded into the Hamiltonian formalism of
statistical mechanics, which we are using throughout this article. 
It is the negative exponent in the formula for the partition function, 
given in eq.\ (\ref{prob}).}
\be  \label{O3Hami}
\frac{1}{T} {\cal H} [\vec e \, ] = \int d^{2} x \ \Big[ \frac{1}{2g} \, 
\partial_{\mu} \vec e \cdot \partial_{\mu} \vec e 
- \frac{\theta}{8 \pi} \ri \, \epsilon_{\mu \nu} \, 
\vec e \cdot (\partial_{\mu} \vec e \times \partial_{\nu} \vec e 
\, ) \Big] = \frac{1}{T} {\cal H}_{0} - \ri \, \theta \, Q[\vec e \, ] \ .
\ee
The 3-component classical spin field $\vec e (x) \in S^{2}$ has the form 
that we wrote down for the Heisenberg model (below eq.\ (\ref{Hami})).
The term ${\cal H}_{0}$ is just a continuum version of the form
(\ref{Hami}) at $\vec H = \vec 0$, up to
the notation for the coupling constant. At large spin $s$, 
the (approximately classical) spin $\vec S$ can be written as
$\vec S \simeq s \, \vec e$, still with the convention $|\vec e\, | 
=1$, which leads to a weak coupling $g \simeq T/(J s^{2})$.

The important novelty is the {\em $\theta$-term:}
its integrated form, $Q [\vec e \, ]$, counts how many times
the configuration  $[\vec e \, ]$ covers the sphere $S^{2}$
in an oriented manner. Hence it is an integer, namely
the {\em topological charge}, or {\em winding number,}
$Q [\vec e \, ] \in \ZZ = \Pi_{2}(S^{2})$.\,\footnote{Small 
variations of a configuration (except for a subset of measure 
zero) do not change $Q [\vec e \, ]$, so the $\theta$-term is 
not visible in the field equations of motion, nor in 
perturbation theory (expansion in powers of $g$). 
Still, it does affect the actual physics, which is
non-perturbative (finite $g$).\label{nonpert}}
Therefore $\exp(-{\cal H}/T)$ is $2\pi$-periodic in $\theta$, 
so it is sufficient to consider $0 \leq \theta < 2\pi$.

The Haldane-Affleck map of an anti-ferromagnetic quantum
spin chain onto this model relates the quantum spin $s$ to
the {\em vacuum angle} $\theta$ as $\theta = 2\pi s$
(within the large $s$ construction) \cite{Haldane2,Affmap,Affrev}. 
Taking into account the $2\pi$-periodicity 
in $\theta$, we infer the scheme

\begin{center}
\begin{tabular}{c cc cc}
 && && Haldane Conjecture \\
$s$ integer && $\theta = 0$ && gap \\
$s$ half-integer &&  $\theta =\pi$ && gapless
\end{tabular} \\
\end{center}

Under this mapping, the Haldane Conjecture takes a new turn.
It is remarkable that the mysterious part flips to the other
side: the gap for the 2d O(3) model without a $\theta$-term
is well established, see {\it e.g.}\ Refs.\ \cite{Polyakov}. 
On the other hand, it is hard to verify whether the limit $\theta = \pi$
is indeed gapless. If the mapping were rigorous, we could conclude 
(considering Ref.\ \cite{AffLieb}) that everything is accomplished, but
it is another conjecture. Hence the challenge is to investigate
the case $\theta = \pi$.

Perturbation theory does not help (cf.\ footnote \ref{nonpert}),
so {\em Ian Affleck} (born 1952 in Vancouver) suggested a 
non-perturbative topological picture \cite{meronpic},
along the lines of Section 2.
Affleck starts from ${\cal H}_{0}$ and adds 
an auxiliary potential term $\sim \mu^{2} (e^{(3)}(x))^{2}$,
which pushes the field $\vec e$ into the $(e^{(1)},e^{(2)})$-plane;
in the limit $\mu^{2}\to \infty$ we are back with the 2d XY model.
We call $\vp$ the angle within this preferred plane (as before), 
and $\alpha$ the (suppressed) angle out of it.

Let us consider a sub-volume, where the configuration contains
a V or an AV in the preferred plane. Its contribution to the topological
charge $Q$ is given by the vorticity computed in eq.\ (\ref{v1}),
using assumption (\ref{nababs}) and Stokes' Theorem, 
but now normalized by the area of $S^{2}$,
\be  \label{meroncharge}
q = \frac{1}{4\pi} \oint d \vec x \cdot \vec \nabla \vp (x) =
\pm \frac{1}{2} \ .
\ee
Local topological defects of this kind, with $q=1/2$ and $q=-1/2$,
are denoted as {\em merons} and {\em anti-merons}, respectively.

The energy estimate is similar to eq.\ (\ref{EV}), in particular
we still obtain the factor $\ln L/a$ (we now write explicitly
a ``lattice spacing'' $a$). The large-$L$ limit only allows for
configurations with total vorticity 0, as before, but it permits
meron--anti-meron pairs (cf.\ eq.\ (\ref{EVAV})). At the end we have 
to remove the auxiliary potential, $\mu^{2} \to 0$; then the merons
and anti-merons
can easily avoid the UV divergence in the core, by choosing 
spin directions out of the previously preferred plane.

Hence we arrive at a picture, which allows for numerous
merons and anti-merons, which hamper the long-range order,
and cause the energy gap. In addition to the meron--anti-meron
pairs, there can be an excess of one type by an even number, such 
that $Q = (n_{\mbox{meron}}-n_{\mbox{anti-meron}})/2 \in \ZZ$.\,\footnote{We 
neglect higher topological defects, which are 
exceptional at weak coupling.}
However, the (not so tight) meron--anti-meron pairs are mainly 
responsible for the energy gap.

So far this is the picture for $\theta =0$. If we now include a vacuum 
angle $\theta$, we see from eqs.\ (\ref{O3Hami}), (\ref{meroncharge})
that this attaches to each region with a meron, or an anti-meron, a 
factor $\exp(\pm \ri \theta /2)$. Therefore any sub-volume with a
meron--anti-meron pair picks up a factor $\cos (\theta /2)$.
In particular $\theta = \pi$ ``neutralizes'' all these
pairs: they do not appear in $\exp(-{\cal H}/T)$, so they cannot 
erase the long-range order anymore, and the gap vanishes.

\begin{figure}[h!]
\begin{center}
\includegraphics[width=0.6\textwidth,angle=0]{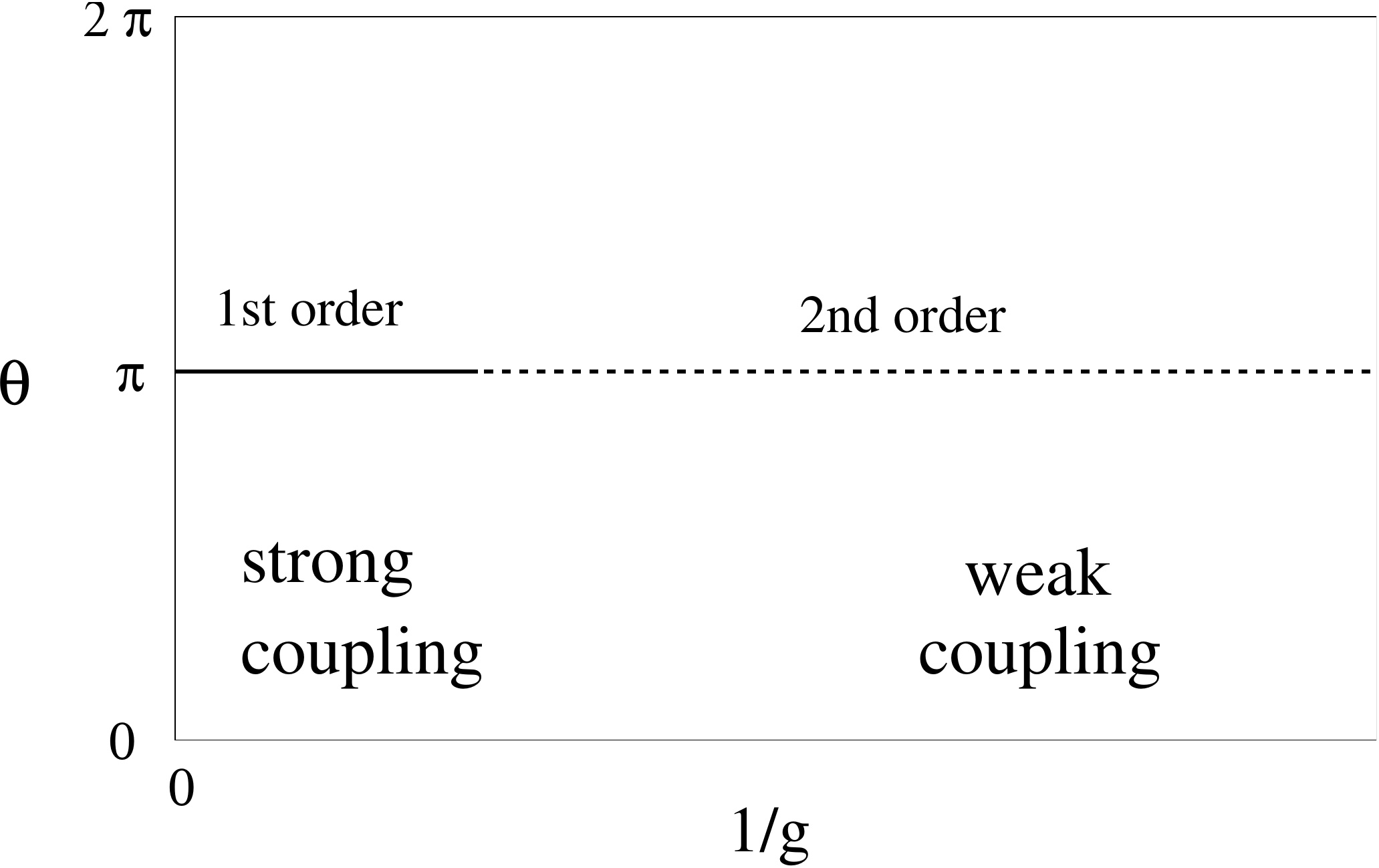}
\caption{The expected phase diagram of the 2d O(3) model with
a topological $\theta$-term. At weak coupling, the map from 
anti-ferromagnetic
quantum spin chains, along with Haldane's Conjecture, predicts a
finite energy gap at $\theta = 0$, but a gapless second order
phase transitions for $\theta \to \pi$.}
\label{phasedia}
\end{center}
\vspace*{-5mm}
\end{figure}

This picture refers to rather smooth configurations, which dominate
at weak coupling, {\it i.e.}\ at small $g$. This is the framework of
the effective low energy theory \cite{Haldane2,Affmap}, and we also 
mentioned that the mapping at large $s$ leads to a small
$g \propto 1/s^{2}$.
For the other extreme, $g \gg 1$, Seiberg reported a cusp in the
free energy, which signals a first order phase transition,
at $\theta = \pi$ \cite{Seiberg}.

Taking these conjectures together, we arrive at the expected
phase diagram shown in Fig.\ \ref{phasedia}. In particular,
if we fix a small (or moderate) $g$, we should
run into a second order phase transition, and therefore into
a continuum limit, for $\theta \to \pi$.

A subtle study in Ref.\ \cite{AGSZ} made this interesting feature
quantitative. To this end, it related the 2d O(3) model at 
$\theta \approx \pi$, at low energy, to a model of conformal 
field theory, known as the $k=1$ Wess-Zumino-Novikov-Witten model 
($k$ is the central charge) \cite{WZNW}, see also Ref.\ \cite{Shank}. 
Assuming both to be in the same universality class (cf.\ Section 1),
the asymptotic behavior of the correlation length was derived as
\be
\xi (\theta \approx \pi) \propto 
\frac{|\ln (|\theta - \pi |)|^{1/2}} {|\theta - \pi|^{2/3}} \ .
\ee
In a finite volume $L \times L$, this translates further into
a finite size scaling of the magnetic susceptibility $\chi_{\rm m}$ 
(given in eq.\ (\ref{gamma})), and the topological susceptibility
$\chi_{\rm t} = (\la Q^{2} \ra - \la Q\ra^{2})/V$: they are both
predicted to exhibit a dominant scaling $\propto L$; for a (more 
abrupt) first order transition one would expect susceptibilities 
$\propto L^{2}$ (generally $L^{d}$).
The conjectured form, refined by logarithmic corrections, reads
\be  \label{gmgt}
\chi_{\rm m} = L \sqrt{\ln L} \, g_{\rm m} (L/\xi ) \ , \quad
\chi_{\rm t} = \frac{L}{\sqrt{\ln L}} \, g_{\rm t} (L/\xi ) \ ,
\ee
where $g_{\rm m}$ and $g_{\rm t}$ are ``universal functions''
with respect to variations of $L$ and $\xi$.

This is an explicit prediction, to be verified in order
to check the above conjecture about a second order phase transitions
for $\theta \to \pi$. The way to study effects
beyond perturbation theory, from first principle, are numerical
Monte Carlo simulations of the lattice regularized model
(we recall footnote \ref{lattice} and Refs.\ \cite{hadphys}).
Its idea is to generate numerous random configurations with
probability $p[\vec e \, ] \propto \exp (- {\cal H}[\vec e \, ]/T)$,
cf.\ eq.\ (\ref{prob}). A large set of such configurations
enables the numerical measurement of expectation values
of the physical terms.

This is straightforward for ${\cal H}_{0}$, but as soon
as we include $\theta \neq 0$, ${\cal H}$ and $\exp(-{\cal H}/T)$
become complex, so they do not define a probability anymore. 
We could generate the configurations using
$|\exp(-{\cal H}[\vec e \, ]/T) |$, and include a
complex phase {\em a posteriori} by re-weighting the statistical
entries. This is correct in principle, but the re-weighting
involves lots of cancellations, hence a reliable measurement
requires a huge statistics --- the required number of 
configurations grows exponentially with the volume $V$.
This is the notorious {\em sign problem.}

In most cases where this problem occurs, in particular in QCD
at high baryon density, and also in QCD with a $\theta$-term,
it has prevented reliable numerical results. However, in the 
case of the 2d O(3) model, this problem was overcome thanks to 
the exceptionally powerful {\em meron cluster algorithm} 
\cite{meroclu}, applied to the constraint Hamilton 
function (\ref{Hcon}) at $\delta = 2 \pi /3$.

This algorithm divides the lattice volume into connected 
sets of spin variables $\vec e_{x}$, the {\em clusters,} 
which are updated collectively \cite{Ulli}. This approach
provides huge statistics (including many configurations that 
do not need to be generated explicitly, 
``improved estimator''). Hence in this
exceptional case, conclusive numerical results were obtained,
and they clearly confirmed the predicted large-$L$ scaling 
of eq.\ (\ref{gmgt}),
including the $\ln L$-refinement \cite{meroclu}.

In addition, the algorithm assigns an integer or half-integer 
topological charge $q$ to each cluster (they sum up to the
topological charge $Q\in \ZZ$ of the entire configuration).
At weak coupling, most clusters are neutral ($q=0$),
and a few percent carry charge $q = \pm 1/2$ (higher
charges are very scarce). At this point, we return to
Affleck's picture, and interpret the clusters with $q=1/2$
($-1/2$) as merons (anti-merons). Then the picture of 
pair neutralization appears in a new light: now it is 
stochastic, and it does not require any O(3) symmetry breaking 
(unlike the potential $\sim \mu^{2} (e^{(3)})^{2}$). Hence it 
confirms the result for the second order phase transition, and it 
even endows the heuristic picture with a neat stochastic 
interpretation.

\begin{figure}[h!]
\begin{center}
\includegraphics[width=0.44\textwidth,angle=0]{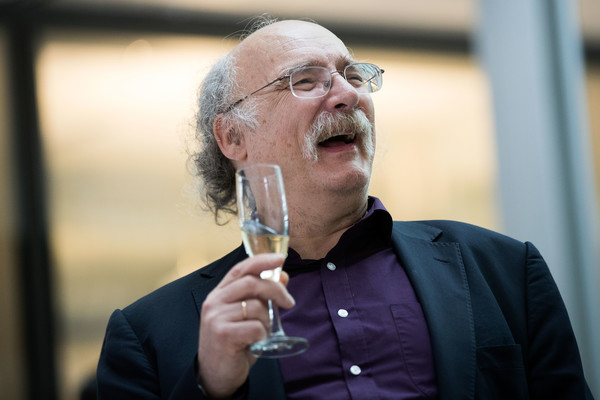}
\hspace*{4mm}
\includegraphics[width=0.4\textwidth,angle=0]{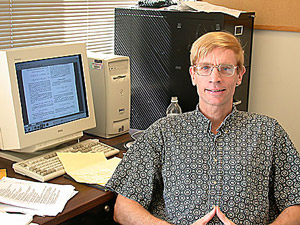}
\caption{{\bf Frederick Duncan Michael Haldane} (left), 
was born 1951 in London and studied at Cambridge University, 
where he graduated in 1978. After working at the University of 
Southern California, the Bell Laboratories and the University 
of California, San Diego,
he became Eugene Higgins Professor at Princeton University in 1990.
{\bf Ian Keith Affleck} (right) was born 1952 in Vancouver, 
studied at Trent University (in Ontario, Canada),
and graduated 1979 at Harvard University,
his Ph.D. advisor was Sidney Coleman. He worked at Princeton University
and Boston University, and since 2003 he is Killam Professor at the
University of British Colombia in Vancouver.} 
\label{bio2}
\end{center}
\vspace*{-5mm}
\end{figure}

\section{Summary}

We described the concept of classical and quantum spin models,
the framework of the 2016 Physics Nobel Prize. We addressed
aspects related to topology, {\it i.e.}\ to quantities which are
invariant under (most) small deformations, and which can only change
in discrete jumps.
We referred to low dimensions, $d=1$ and $2$, where --- at finite 
$T$ --- thermal fluctuations 
prevent the spontaneous breaking of continuous symmetries \cite{MW}, 
and therefore the dominance of ordered structures, 
but higher order phase transitions happen nevertheless.

In the classical 2d XY model we described the BKT phase 
transition \cite{KT}, which is essential (of infinite order), 
and driven by the (un-)binding of vortex--anti-vortex pairs. 
This transition has been observed experimentally, for instance 
in superfluids \cite{superfluid} and in superconductors 
\cite{supercon}, and recently also in systems of ultra-cold 
atoms trapped in optical lattices \cite{coldatoms,Kor}.

Then we summarized the history of anti-ferromagnetic quantum spin 
chain studies, in particular the Haldane Conjecture 
\cite{Haldane1,Haldane2} about 
energy gaps for integer spin vs.\ gapless chains for 
half-integer spin. This insight agrees with experimental
results as well \cite{ex1,ex2}. We further discussed the mapping onto
a classical 2d O(3) model with a topological $\theta$-term
(the Haldane-Affleck map \cite{Haldane2,Affmap}), 
and the manifestation of the Haldane Conjecture in that system.

These are only selected topics of the works, which were
awarded with the Physics Nobel Prize 2016.
For a review of aspects which have not been covered here --- 
in particular the quantum Hall effect and topological 
insulators --- we refer to Ref.\ \cite{Melo}.

\ \vspace*{5mm} \\
{\bf Acknowledgements:} 
We thank Michael B\"{o}gli, Ferenc Niedermayer, Michele Pepe, Andrei 
Pochinsky, Fernando G.\ Rej\'{o}n-Barrera and Uwe-Jens Wiese for their 
collaboration in projects related to the BKT transition and the 
Haldane conjecture, and Hosho Katsura for a helpful remark
regarding Ref.\ \cite{AKHT}.

This work was supported by the Albert Einstein Center for
Theoretical Physics, the  European  Research  Council under the  
European  Union's  Seventh  Framework Programme 
(FP7/2007-2013)/ERC grant agreement 339220, the Consejo Nacional de 
Ciencia y Tecnolog\'{\i}a (CONACYT) through project CB-2013/222812, 
and by DGAPA-UNAM, grant IN107915.

\end{document}